%% file: Barrado_B30_Euclid.tex
\documentclass[]{aa} 

\usepackage{graphicx}
\usepackage{lscape}
\usepackage{longtable}
\usepackage{graphicx}
\usepackage{times}
\usepackage{graphicx}
\usepackage{xspace}
\usepackage{epsfig}
\usepackage{natbib}
\usepackage{rotating}
\usepackage{dcolumn}
\usepackage{xcolor}
\usepackage{enumerate}
\usepackage{lineno}
\linenumbers
\usepackage{txfonts}
\newcommand{\sfont}[1]{{\scriptscriptstyle\rm #1}}
\newcommand{\IE}{\ensuremath{I_\sfont{E}}\,}
\newcommand{\YE}{\ensuremath{Y_\sfont{E}}\,}
\newcommand{\JE}{\ensuremath{J_\sfont{E}}\,}
\newcommand{\HE}{\ensuremath{H_\sfont{E}}\,}

\begin{document}


\title{Euclid Early Release Observations of the Barnard 30 dark cloud}

\subtitle{I. Brown dwarfs and planetary mass candidate members at the core of the association}

   \author{D. Barrado
          \inst{1}
          \and
          H. Bouy
          \inst{2,3}
          \and
          E. L. Martin
          \inst{4}
          \and
          J.-C. Cuillandre
          \inst{5}
          N. Hu\'elamo
          \inst{1}
          \and
          M. \v{Z}erjal
          \inst{4,6}
          \and
          C. Dominguez-Tagle
          \inst{4,6}
          \and
          S. Mu\~noz Torres
          \inst{4,6}
          \and
          J.-Y. Zhang
          \inst{4,6}
          \and
          E. Bertin
          \inst{5}
          \and
          J. Olivares 
          \inst{7}
   }

   \institute{Centro de Astrobiolog\'{\i}a (CAB), CSIC-INTA, ESAC Campus, Camino bajo del Castillo s/n, E-28692 Villanueva de la Ca\~nada,
Madrid, Spain\\
              \email{barrado@cab.inta-csic.es}
         \and
         Laboratoire d’astrophysique de Bordeaux, Univ. Bordeaux, CNRS, B18N, allée Geoffroy Saint-Hilaire, 33615 Pessac, France
         \and
         Institut universitaire de France (IUF), 1 rue Descartes, 75231 Paris CEDEX 05, France
         \and
         Instituto de Astrof\'{\i}sica de Canarias, 38205 La Laguna, Tenerife, Spain
         \and
         Université Paris-Saclay, Universit\'e Paris Cit\'e, CEA, CNRS, AIM, 91191, Gif-sur-Yvette, France
         \and
         Universidad de La Laguna, Dpto. Astrof\'{\i}sica, 38206 La Laguna, Tenerife, Spain
         \and
         Departamento de Inteligencia Artificial, Universidad Nacional de Educación a Distancia (UNED), c/Juan del Rosal 16, E-28040, Madrid, Spain
   }

   \date{Received ; accepted }

 
  \abstract
   {}
{This study aims to identify very low-mass members within the Barnard 30 dark cloud, with a particular focus on detecting objects in the substellar domain, including those within the planetary-mass regime.}
{We employed deep photometric observations using data from the \textit{Euclid} mission, incorporating advanced data processing techniques—most notably, the DeNeb tool—for optimized source detection. We also analyze multi-wavelength ancillary observations and perform a Spectral Energy Distribution analysis for each candidate member. In addition, low-resolution near-infrared (NIR) spectroscopy was obtained for one candidate to further assess its nature and properties.}
{Our initial photometric analysis yielded a initial sample of nearly one hundred candidate members in the substellar mass range. 
A subsample of 23 probable members, located close to the 3 Myr isochrone, has been identified. Thus, we have   substantially expanded the known population of faint, cool sources associated with the region. Low resolution near-IR spectroscopic analysis of one candidate reveals an L2 spectral type with low gravity, consistent with a young ultra-cool dwarf. If its membership is confirmed, its estimated mass lies in the range $15$--$20~M_{\mathrm{Jup}}$. These findings validate the reliability of our multi-wavelength photometric selection methodology.}
{These results offer valuable insights into the low-mass end of the initial mass function (IMF) and demonstrate the effectiveness of \textit{Euclid} in identifying brown dwarf and planetary-mass candidates in nearby, densely packed star-forming regions.}






   \maketitle
%
\nolinenumbers

\section{Introduction}

The Lambda Orionis star-forming region \citep{Murdin77.1}, a complex structure located approximately 400\,pc away, contains several distinct associations. These include Collinder 69, the central cluster associated with the $\lambda$ Orionis binary star—an O8 III star \citep{Bouy2009-LOri}—as well as the dark clouds Barnard 35 and Barnard 30 (hereafter B35 and B30).
B30 is situated at the edge of a very large, quasi-circular structure. The evolutionary history of the entire region may have been shaped by a supernova explosion that occurred a few million years ago \citep{Maddalena1987-LOSFR, Cunha1996-LOSFR-SN, Dolan2002.1}.

The area surrounding Lambda Orionis has been the target of numerous surveys aimed at identifying solar-like and low-mass stars \citep{Dolan1999.1, Dolan2001.1, Dolan2002.1, Koenig2015_C69_SOri_YSO_WISE}. In several cases, {\it bona fide} members have been confirmed through spectroscopic analysis of lithium content and radial velocity.

The evolutionary status of this complex remains somewhat uncertain. Collinder 69, the central cluster, is estimated to be approximately 5–8 Myr old \citep{Barrado2004.3, Bayo2011.1}, while the dark clouds B30 and B35 appear to be around 3 Myr old or even younger.

We have dedicated considerable effort to understanding this intriguing and complex region. Our investigations have focused on specific aspects such as circumstellar disks, accretion, and binarity at the low-mass end of the spectrum, well within the substellar regime \citep{Barrado2007.1, Barrado2007.2, Morales2008-PhD, Barrado2011.1, Bayo2011.1, Bayo2012.1}. A particular emphasis has been placed on identifying young objects in the B30 dark cloud, the youngest star-forming region in the complex. We have conducted two submillimeter surveys and identified several embedded sources. Some of these sources are extremely faint and could represent pre- or proto-brown dwarfs—the potential progenitors of brown dwarfs \citep{Huelamo2017-ALMA-B30, Barrado2018_B30}. For a review, see \citep{Palau2024}.

In this work, we present the results of an \textit{Euclid} Early Release Observation (ERO) of the B30 dark cloud. These data were collected as part of a broader initiative to identify very low-mass members across several nearby stellar associations. Further details are available in \citet{Martin2025_Euclid_SOri}. 
The goal is to investigate the power of Euclid to uncover the substellar population of B30 down to the planetary-mass regime, to explore the lower limit of the star formation process, and to relate these properties to the environmental conditions in which they arise.


\section{Analysis}
\label{Analysis}

\subsection{The data}
\label{data}

\subsubsection{Euclid data}
\label{EUCLID_data}


Our B30 field, centered at RA = 05:31:29 (82:52:47) and DEC = +12:20:01, was observed with the \textit{Euclid} mission \citep{Euclid2024_Mission} as part of the Early Release Observations (ERO) program on 2023-10-12. The standard \textit{Euclid} observational strategy was employed, consisting of four dithered exposures of 89 and 87 seconds for the VIS \citep{Euclid2025-VIS} and NISP \citep{Euclid2025_NISP} instruments, respectively. For this study, we utilized the stacked mosaics in all available filters: $I_E$, a single broad band filter from 550\,nm to 950\,nm used with the VIS instrument, and $Y_E$, $J_E$, $H_E$, the three infrared filters used  with the NISP instrument \citep{Euclid2022-NISP}. The data processing methodology is detailed in \citet{Cuillandre2025-Euclid}.

Since B30 is immersed in a dense and inhomogeneous cloud, our \textit{Euclid} image of Barnard 30 was post-processed using DeNeb, a new deep-learning-based tool developed to remove nebular and extended emission from astronomical images (Bertin et al., in prep.). This processing step enhances source detection and improves photometric accuracy, particularly for faint objects embedded in extended emission.

For the near-infrared images taken with the NISP instrument, source detection and photometric extraction were performed using the SExtractor package \citep{Bertin1996-Sextractor}, applying a $\chi^2$ image constructed from a linear combination of the \YE, \JE, and \HE bands. This method follows the approach of \citet{Szalay1999} and is optimized for simultaneous multi-band detection of faint sources.

Photometry in the \IE band from the VIS instrument was extracted for all sources with more than three contiguous pixels above 1.5 times the local background standard deviation. Source positions were determined using the PSF and Sérsic model-fitting option in SExtractor, which utilizes an empirical PSF model previously derived using PSFEx \citep{Bertin2011-SExtractor}.

Our initial catalog includes 525,651 sources with photometry in at least one of the \IE, \YE, \JE, or \HE bands. It is important to note that these filters differ significantly (e.g., \citealt{Euclid2022-NISP, Euclid2025-VIS}) from more conventional photometric systems—particularly the \IE band, which spans a broad spectral range
\footnote{More information about different photometric systems at http://svo2.cab.inta-csic.es/theory/fps/ (\citealt{Rodrigo2024-Filters})}. Further details on data extraction and the initial photometric selection are provided in \citet{Bouy2025-EUCLID} and \citet{Martin2025_Euclid_SOri}.

Our first step was to remove sources with saturated photometry (\IE$<$17.75, mag, \YE$<$16.0 mag, \JE$<$16.0 mag, \HE$<$16.0 mag) and extended objects. Then, following  \citet{Bouy2025-EUCLID}, we excluded any object meeting one or more of the following morphometric criteria:
FWHM(\IE)$<$1.3,  FWHM(\IE)$>$2.1 or abs({SPREAD MODEL \IE})$>$0.003; 
FWHM(\YE)$<$1.2,  FWHM(\YE)$>$2.2 or abs({SPREAD MODEL \YE})$>$0.005; 
FWHM(\JE)$<$1.1,  FWHM(\JE)$>$2.0 or abs({SPREAD MODEL \JE})$>$0.005; 
FWHM(\HE)$<$1.2,  FWHM(\HE)$>$2.2 or abs({SPREAD MODEL \HE})$>$0.005.
Finally, we refined our sample by eliminating  261 nearby and fast moving objects ($\Pi$$>$5 mas and/or $\mu$ $>$6 mas/yr) using Gaia DR3 data  (\citealt{Gaia2016-Mission, Gaia2016-DR1, Gaia2021-DR3-Content}), specially Ultra Cool Dwarfs (UCDs) in the field of B30.
We were left with a sample of 24,304 objects.

As described in \citet{Bouy2025-EUCLID}, these morphometric criteria are effective in removing blended visual binaries that remain unresolved by SExtractor—typically those with separations near the diffraction limit of the VIS instrument (i.e., approximately 0.2 arc-seconds)—as well as young stellar objects exhibiting extended emission. These later cases will be addressed in a separate analysis (Huélamo et al., in prep.).

\subsubsection{Ancillary photometric observations}
\label{ancillary}

Over the past two decades, we have compiled a substantial dataset in the Lambda Orionis star-forming region.  
The datasets related to B30 have been already published
\citep{Huelamo2017-ALMA-B30, Barrado2018_B30},
and contain data from wide-field imaging observations from the following telescopes and instruments:

\begin{itemize}
\item CFHT/MegaCam: $i$ band
\item CTIO/DECam: $VR$, $i$, $z$, and $y$ bands
\item Subaru/HSC: $r$ band
\item UKIRT/WFCAM: $J$, $H$, and $K_s$ bands
\item INT/WFC: $i$ and $Z$ bands
\item CAHA 3.5m/O2000: $J$, $H$, and $K_s$ bands
\end{itemize}

In addition, complementary photometry has been obtained within the COSMIC-DANCE survey \citep{Bouy2013-Dance} and from public archives.

We have also reprocessed archival Spitzer/IRAC images and extracted the photometry following the same methodology described in \citet{Bouy2025-EUCLID}. As a result of the new processing, our Spitzer photometry reaches significantly deeper than that of \citet{Morales2008-PhD}, owing to the combination of all available epochs and the application of DeNeb to suppress bright nebular emission. This results in excellent overlap with the \textit{Euclid} dataset, extending down to the planetary-mass regime.

Altogether, we have assembled a comprehensive multi-wavelength database with coverage in 12 photometric bands, spanning from the optical $r$ band to 4.5 $\mu$m in the mid-infrared.

\subsubsection{Spectroscopy}
\label{SpectroscopicData}

We have obtained a near-infrared spectrum of B30-Euclid-025, one of our candidates, in the wavelength interval of 14540-24050 \AA{}, using the Espectrografo Multiobjeto Infra-Rojo \citep[EMIR, ][]{Garzon2022-EMIR}, mounted at the Nasmyth-A focus of the 10.4-m Gran Telescopio Canarias (GTC) at the Roque de los Muchachos Observatory on the Spanish island of La Palma, as part of the program GTC80-24B (PI: E. Martín). The observation was taken during the night of February 16th 2025, between 21:15 and 23:15 UTC, under a clear dark sky with a 0.7\arcsec seeing. We used the HK low-resolution grism with a 1\arcsec slit, corresponding to a spectral resolution of $R\approx600$. We set a single integration time of 160s and three ABBA cycles with a throw of 6” along the slit to remove the sky contribution, yielding a total on-source integration time of 1920\,s. We also observed a telluric standard star, HD\,283677 (B8; \citealt{Nesterov1995}), at the beginning of the observation with a similar airmass, using a 6\,s single integration time and one ABBA pattern with a throw of 10\arcsec.  

In order to obtain the stacked 2D spectrum, we utilized the EMIR pipeline (PyEmir\footnote{https://pyemir.readthedocs.io/en/latest/index.html}). The pipeline was designed to rectify the spectra, calibrate the wavelength in vacuum using OH air-glow lines, subtract the sky background using the subsequent A/B image, and combine all the sky-subtracted images. Subsequently, we obtained the 1D spectrum employing the \textit{apall} task from the Image Reduction and Analysis Facility \citep[IRAF, ][]{Tody1986-IRAF, Tody1993-IRAF}. To reduce the telluric standard star, the same process was followed, with the addition of masking all the prominent Paschen and Brackett lines in the spectrum using the IRAF \textit{splot} task. Finally, the flux correction and telluric correction were applied simultaneously by multiplying it by a B8V standard spectrum from the ESO library\footnote{https://www.eso.org/sci/observing/tools/standards/IR$\_$spectral$\_$library.html} \citep{Pickles1998} and dividing it by the masked telluric standard spectrum.  The final spectrum was measured to have an average signal-to-noise ratio (SNR) of 20-30 per resolution element.

\section{Membership selection}
\label{Membership}

\subsection{Color-Magnitude Diagrams}
\label{CMD}

\begin{figure}
   \centering
   \includegraphics[width=8.8cm]{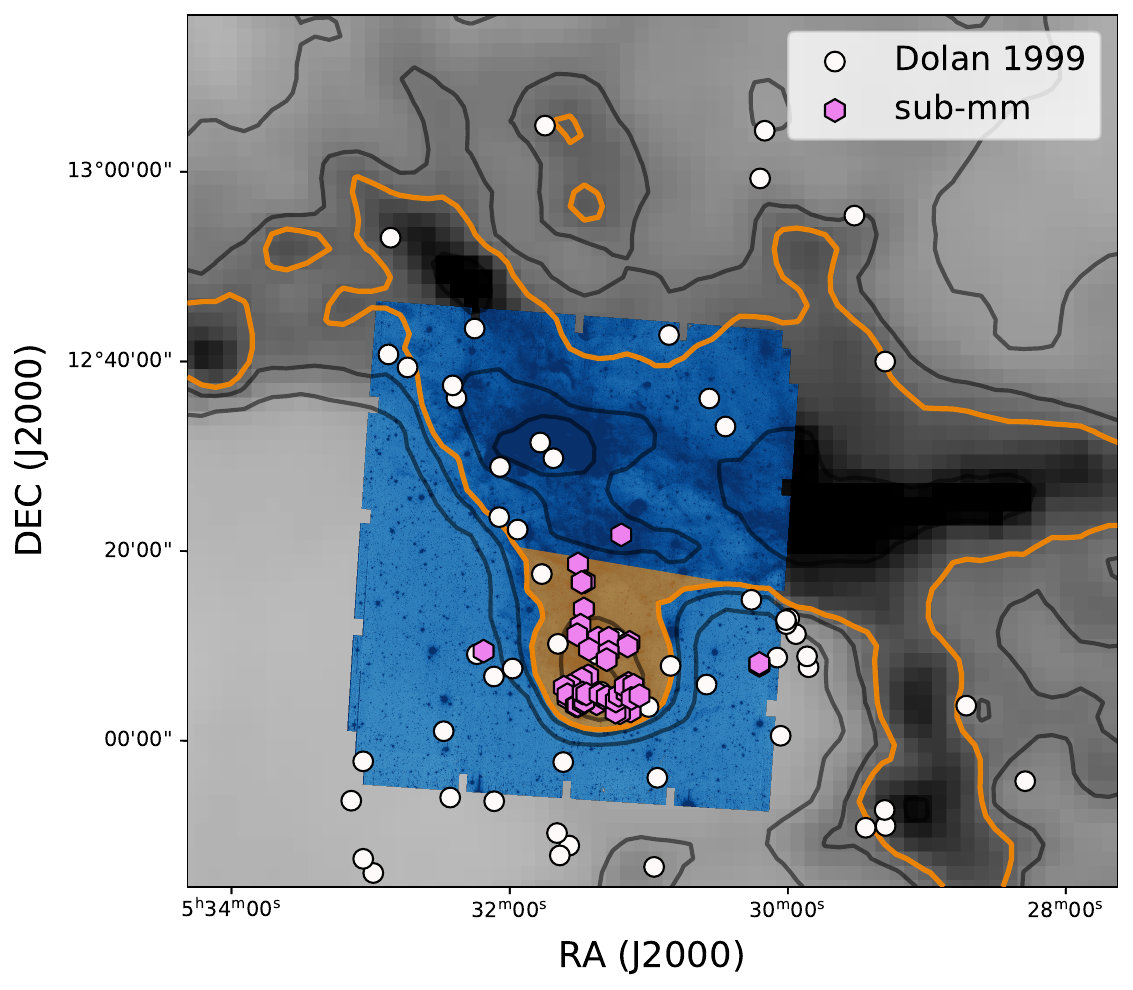}
   \caption{IRAS 100\,$\mu$m negative image of B30. The overlaid blue region corresponds to the observed Euclid ERO field. The filled orange area, defined using the IRAS isocontours, represents the region where we have searched for faint members in B30 using the thick cloud as a shield.
    We also display solar-like members identified through radial velocity selection from \citet{Dolan2001.1} with white circles, and sub-millimeter detections from \citet{Huelamo2017-ALMA-B30} and \citet{Barrado2018_B30} with violet hexagons.}
         \label{F_RA_DEC_IRAS_new}
   \end{figure}
   
\begin{figure}
   \centering
   \includegraphics[width=8.8cm]{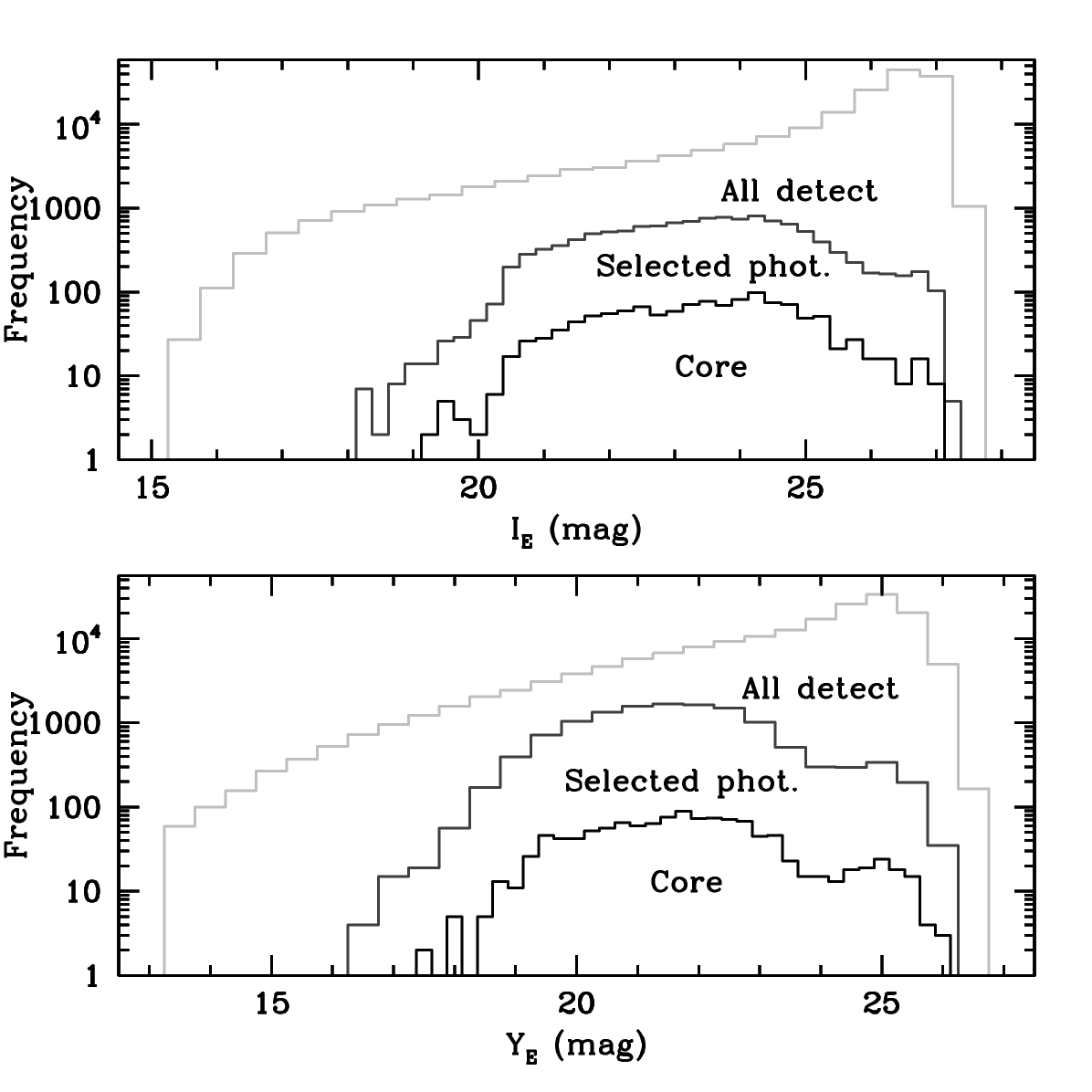}
   \includegraphics[width=8.8cm]{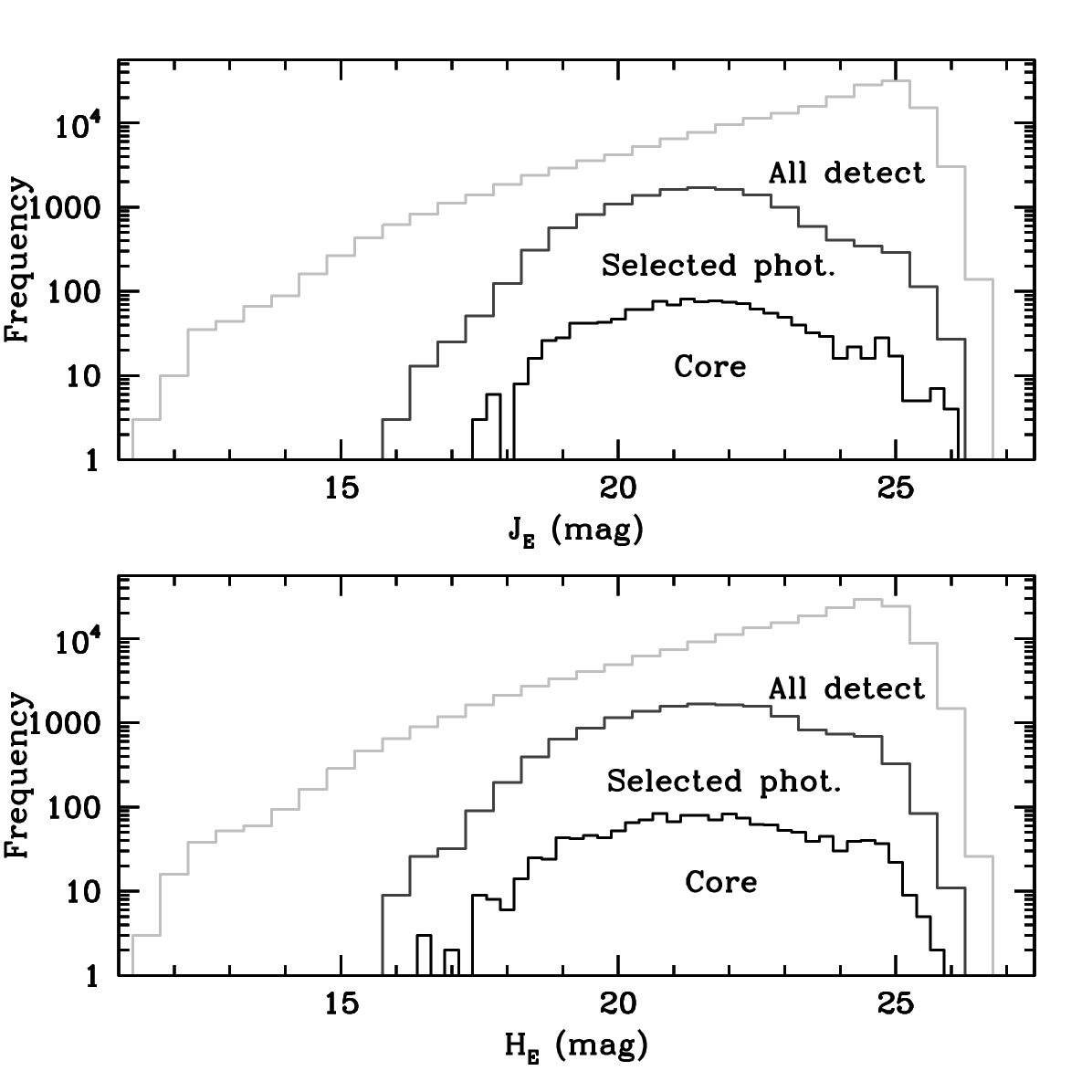}
   \caption{
     Histogram of the number of detections across different samples: Light grey represents the complete dataset (with photometric errors less than 0.5 mag), dark grey shows the selected photometry, and black corresponds to data located within the core of the B30 dark cloud.
   }
         \label{F_Histo_JY}
   \end{figure}

We restricted our search for faint members of B30 to the core of the dark cloud, as defined by the IRAS submillimeter isocontours (Fig.~\ref{F_RA_DEC_IRAS_new}). 
This spatial constraint (i.e., using the thick cloud as a shield) effectively removed a significant fraction of background sources, particularly unresolved extra-galactic contaminants not eliminated during the initial data curation (see Subsection~\ref{EUCLID_data}). Applying this boundary reduced our initial sample to 2,247 sources. A histogram for each \textit{Euclid} band can be found in Fig.~\ref{F_Histo_JY}, were we represent three different samples for each of them: all detections, the selected data once the \textit{Euclid} morphometric criteria  have been applied, and the spatial restriction (i.e., core region). We then identified candidate members using two distinct, but complementary, selection strategies.

   \begin{figure*}
   \centering
   \includegraphics[width=8.8cm]{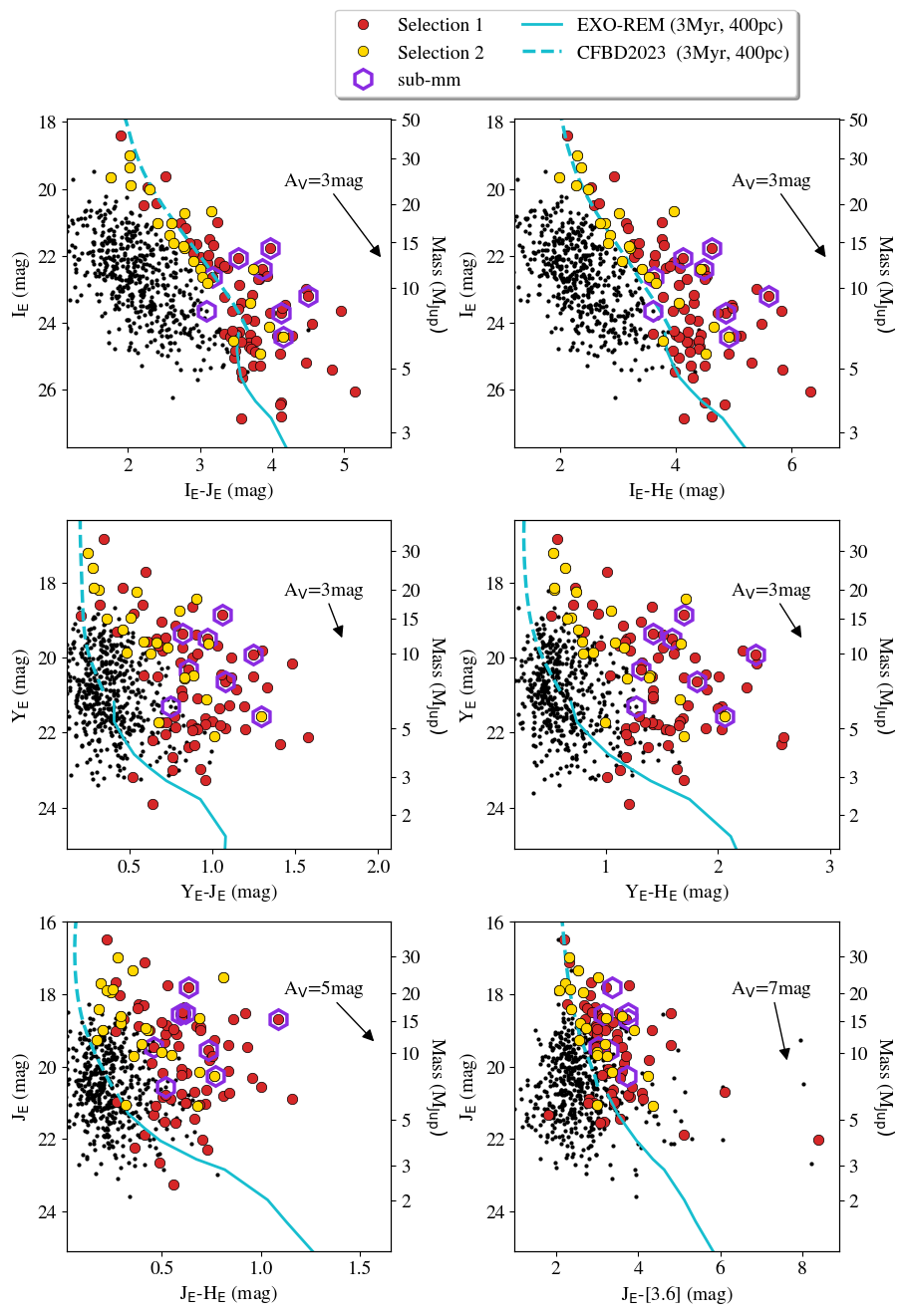}
   \caption{
     Euclid color-magnitude diagrams. The initial B30 sample is shown as gray dots, while our candidate members are indicated by red (selection \#1) and orange (selection \#2) circles. Previously known B30 candidates from \citet{Huelamo2017-ALMA-B30} and \citet{Barrado2018_B30} are marked with purple hexagons. The 3 Myr isochrones at 400 pc corresponding to CFBD2023 \citep{Chabrier2023} and EXO-REM \citep{Charnay2018-BD} models are plotted with dashed and solid light blue curves, respectively. The corresponding mass scale is shown on the right-hand axis. Extinction vectors are also included.
   }
         \label{F_CM_Euclid}
   \end{figure*}

The first selection method is the most conservative. We constructed six color–magnitude diagrams (CMDs; Fig.~\ref{F_CM_Euclid}) using the \textit{Euclid} filters, supplemented with the Spitzer [3.6] band. We employed the 3 Myr theoretical isochrones from \citet[][CFBD2023]{Chabrier2023}, computed specifically for the \textit{Euclid} and Spitzer filter systems, down to effective temperatures of 2000 K. 
Below 1800 K, the CFBD2023 models fail to reproduce the observed colors and luminosities of young late-type objects. We therefore used the CFBD2023 isochrones down to 1800 K, and complemented them below this threshold with 3 Myr isochrones computed from the EXO-REM model grid assuming solar metallicity and cloud properties as \verb|f_sed|=3 \citep[see][]{Charnay2018-BD}.

Candidate members were then selected following the procedure outlined in \citet{Bouy2025-EUCLID}. Specifically, a source was considered a candidate only if it fell on the red side of the isochrone in all CMDs, after accounting for photometric uncertainties. A total of 71 objects met these strict criteria and are marked as red circles in Fig.~\ref{F_CM_Euclid}.

We expect minimal contamination from foreground stars and background galaxies in this sample. 
For the former, the density of nearby stars with similar photometric properties to {\it bona fide} B30 members is low (i.e., closer ultra-cool dwarfs  at just the right distance to blend in with members of the association). 
A fraction of these objects was already removed from our initial catalog using Gaia data. In any case, the Euclid substellar candidates are in fact redder and fainter than Gaia detection limits. 
For the latter, extended sources (i.e., galaxies) were removed during our initial filtering, and the presence of bright star-forming nebulosity serves as a natural barrier against distant background sources, specially point-like ones not removed previously. The selected candidates, along with their \textit{Euclid} photometry, are presented in Tab.~\ref{TAB_EuclidCanSelHB}. Among these 71 candidate members, seven correspond to previously identified B30 objects reported by \citet{Huelamo2017-ALMA-B30, Barrado2018_B30} through the analysis of their SEDs, including sub-mm observations: B30-LB03i, B30-LB05c, B30-LB17b, B30-LB18e, B30-LB27a, B30-LB28a, and B30-LB30e. Contamination by reddened background stars might be relevant. 
However, note that the reddening vector of different CMD in Fig.~\ref{F_CM_Euclid} run almost parallel to the isochrone or the cluster sequence. Therefore, the main effect should be on the mass estimate for cluster members and contamination by stellar or extra-galactic sources should be limited.

       \begin{figure}
       \centering
       \includegraphics[width=8.8cm]{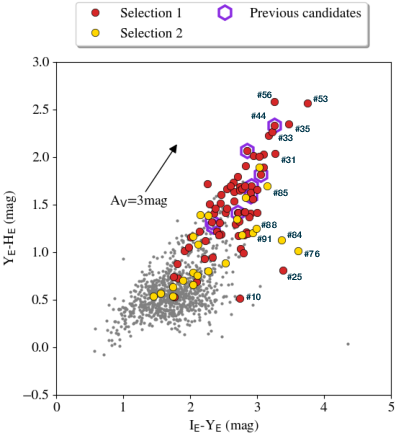}
       \caption{
         Euclid color-color diagram. Some candidates might have a  strong reddening.
         Symbols as in Fig.~\ref{F_CM_Euclid}. Labels correspond to the object ID listed in the tables.
       }
             \label{F_CC}
       \end{figure}

Our second selection method employs more flexible criteria, taking advantage of our multi-wavelength dataset, which includes optical, ground-based near-infrared, and mid-infrared photometry (Fig.~\ref{F_CM_OptIR}). Given the uncertainties in the models 
and the photometric errors in different bands, we have enlarged the area in the color-magnitude diagrams, selecting candidates located in a wider range around the fiducial isochrone, and including a band in the blue side (by offsetting downwards the isochrone by 0.2 magnitudes). The goal is to include in our selection process all possible candidates, even including a significant number of contaminants.
This process yielded additional 23 candidate members, shown as orange circles in Fig.~\ref{F_CM_Euclid}. These additional candidates are listed  in  Tab.~\ref{TAB_EuclidCanSel2}. 

As a sanity check, we have included in the CMD panels ten previously known B30 candidate members  from  
\citet{Huelamo2017-ALMA-B30} and \citet{Barrado2018_B30}. They were identified based on their sub-millimeter properties. 
Two of them exhibit incomplete photometric coverage and therefore do not satisfy our stricter photometric selection criteria. Nonetheless, they meet the basic quality requirements outlined in Subsection~\ref{EUCLID_data} (e.g., not flagged as having poor photometric quality or extended morphology), so they are also considered as candidates. 
The eight objects (out of the ten) with complete Euclid photometric data are highlighted with magenta hexagons in Fig.~\ref{F_CM_Euclid}.

In summary, we have assembled a comprehensive dataset of B30 candidate members by merging all three previously described samples: 71 candidates from the first selection method, 23 candidates from the second selection method, and 10 candidates from previous sub-mm studies meeting the quality requirements described in Subsection~\ref{EUCLID_data}. This results in a final list of 104 low-mass, possible members of this ~3 Myr-old dark cloud.

   \subsection{Reddening based on the Euclid CCD}
   \label{CCD}

   Fig.~\ref{F_CC} displays a color-color diagram (CCD) with our selection of potential members of the B30 dark cloud.
   A strong reddening should be present in a significant number of our candidate members. Thus, they  should appear significantly fainter than they actually are. Note, however, that there is a significant uncertainty in the locus of the unreddened cluster sequence, specially at the lower end.

      Four sources are of particular interest, namely B30-Euclid-76, 84, 85 and 91, located at the faintest end in several Color-Magnitude diagrams, and B30-Euclid-31, 33, 35, 53 and 56, the reddest objects in our sample. B30-Euclid-44, aka LB03i, is also in this group. Their location in the CC and CM diagrams makes them excellent candidates as {\it bona fide} substellar members of the association, although only a follow-up spectroscopic program would unveil the nature of these candidates.

\subsection{Binarity}
\label{Binarity}

  We have identified several pairs of potential wide binaries -—both components are possible members of the association—- with separations less than 5 arc-seconds (or 2000\,au at a distance of 400\,pc):
 B30-Euclid-08 (J053114.02+120327.8, from selection 1), aka LB30-LB30e, and B30-Euclid-77 (J053114.02+120325.7, from selection 2) at 2\farcs029 ($\sim$812\,au);
 B30-Euclid-44 (J053130.60+121734.6, from selection 1), aka LB30-LB03i, and B30-Euclid-45 (J053130.79+121736.3, also from selection 1) at 3\farcs293 ($\sim$1317\,au);
 B30-Euclid-56 (J053132.01+121308.6, from selection 1) and B30-Euclid-53 (J053131.70+121306.9, also from selection 1) at 4\farcs809 $\sim$1924\,au); and
 B30-Euclid-55 (J053132.01+120408.3, from selection 1) --possible a non-member, see subsection~\ref{HRD}--   and the submillimeter sources B30-LB22c \& B30-LB22j at 3\farcs374 ($\sim$1350\,au) and 4\farcs920 ($\sim$1968\,au), respectively.

\begin{figure*}
      \centering
      \includegraphics[width=8.8cm]{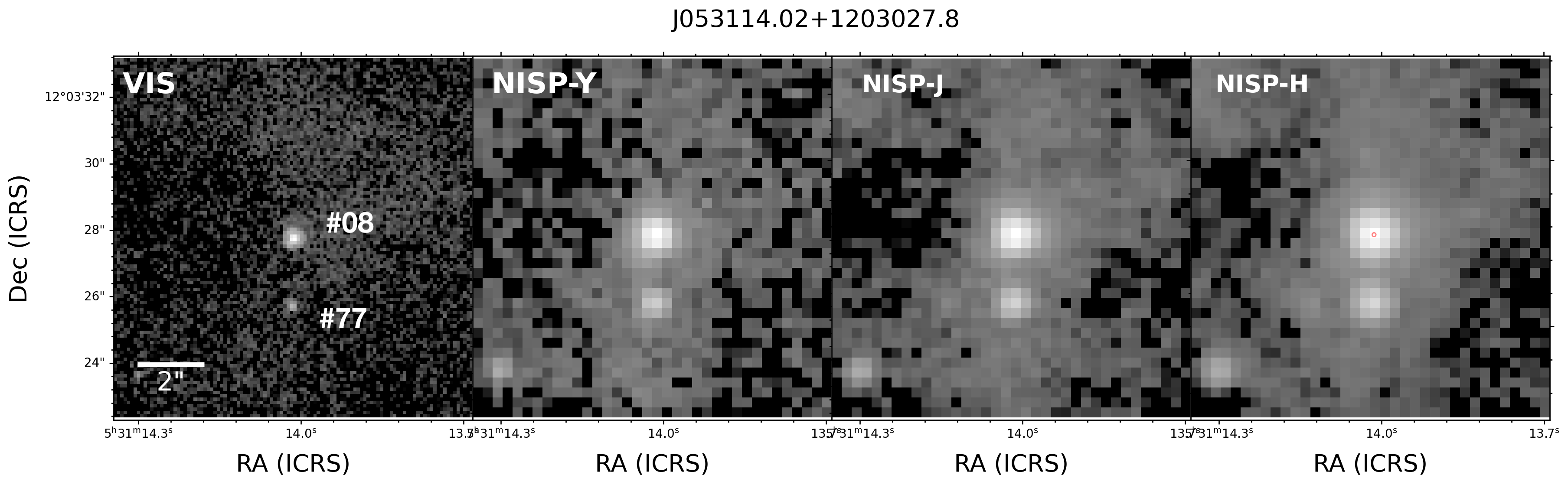}
      \includegraphics[width=8.8cm]{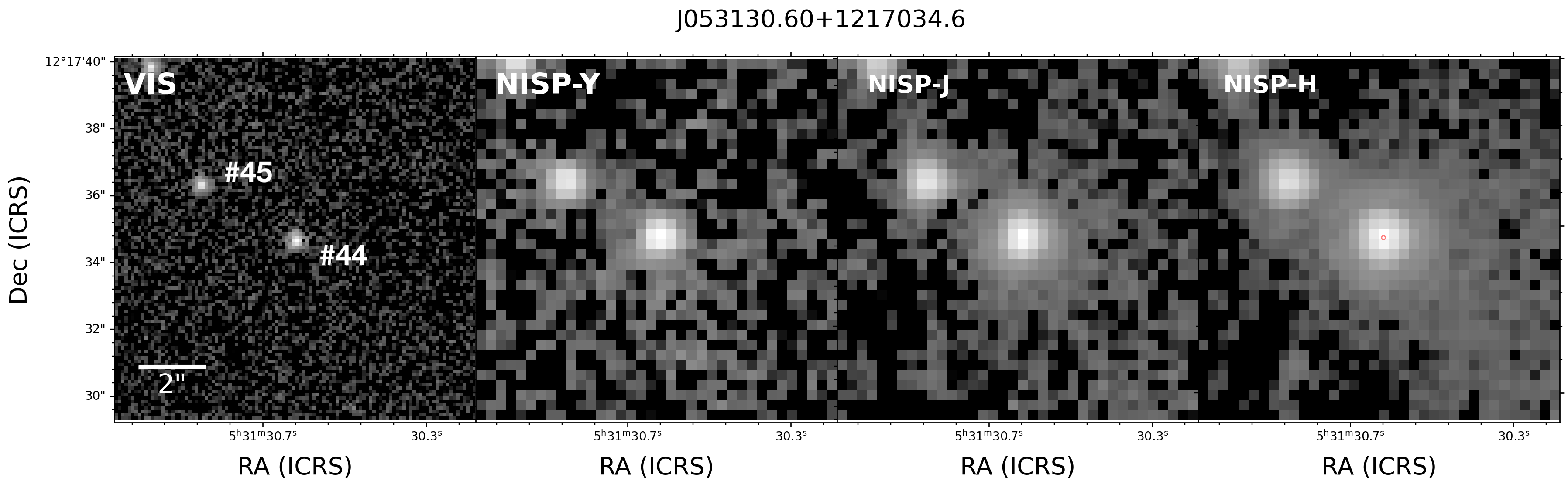}
      \includegraphics[width=8.8cm]{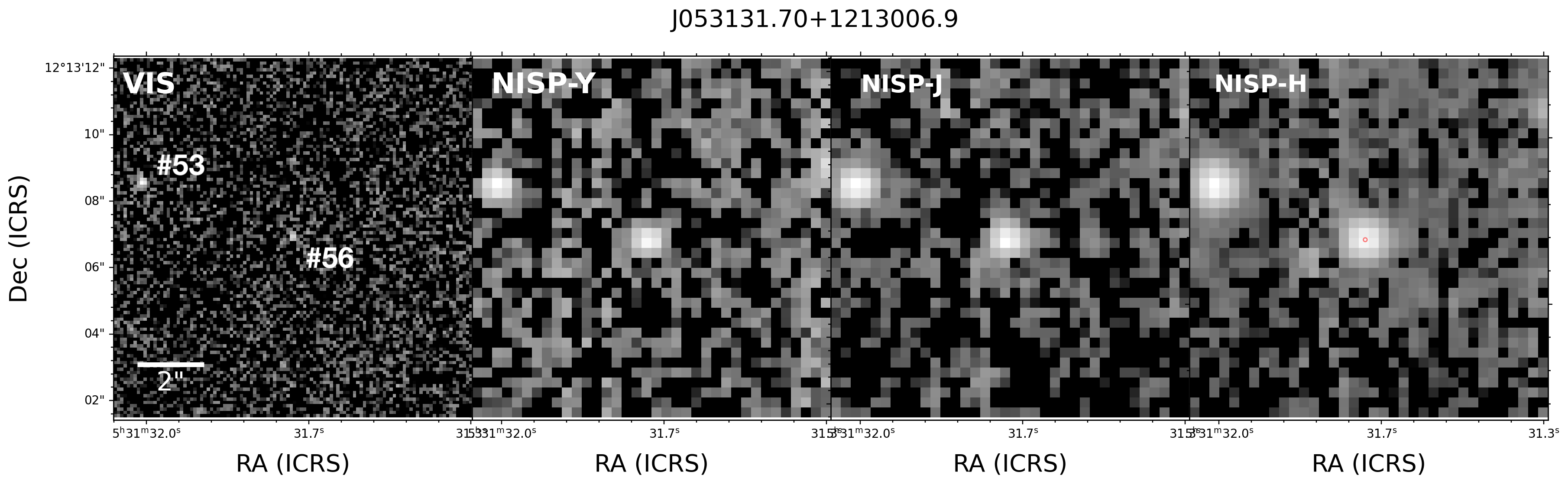}
      \includegraphics[width=8.8cm]{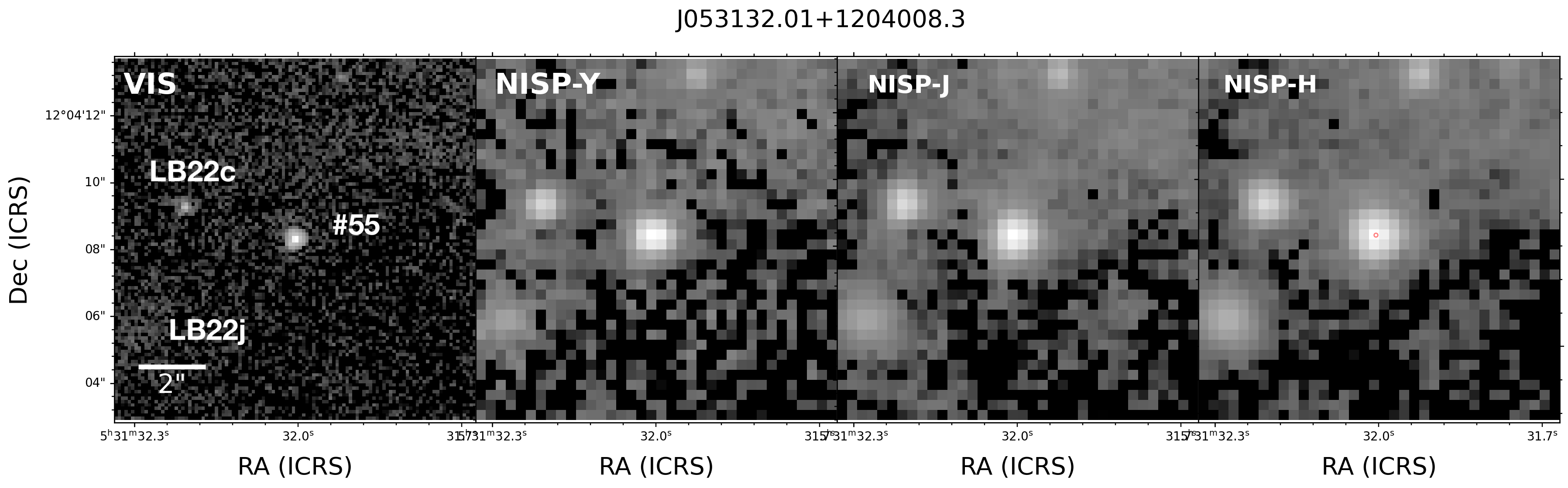}
      \caption{
        Binaries in the B30 dark cloud. All components are candidate members of the young stellar association.
 B30-Euclid-08 (J053114.02+120327.8), aka LB30-LB30e, and B30-Euclid-77 (J053114.02+120325.7) at 2.029";
 B30-Euclid-44 (J053130.60+121734.6), aka LB30-LB03i, and B30-Euclid-45 (J053130.79+121736.3) at 3.293";
 B30-Euclid-56 (J053132.01+121308.6) and B30-Euclid-53 (J053131.70+1213006.9) at 4.809"; and
 B30-Euclid-55 (J053132.01+120408.3) and the sub-mm sources B30-LB22c \& B30-LB22j at 3.374 and 4.920", respectively.}
            \label{F_FChart_Bin}
      \end{figure*}

Since the total sample includes 106 unique sources, the binary fraction is below 2\%, although this estimate may be biased by our selection process. Finding charts in all four \textit{Euclid} filters are provided in Fig.~\ref{F_FChart_Bin}.

\subsection{VOSA and the HRD: SEDs, Teff and Luminosities}
\label{HRD}

   \begin{figure}
   \centering
   \includegraphics[width=8.8cm]{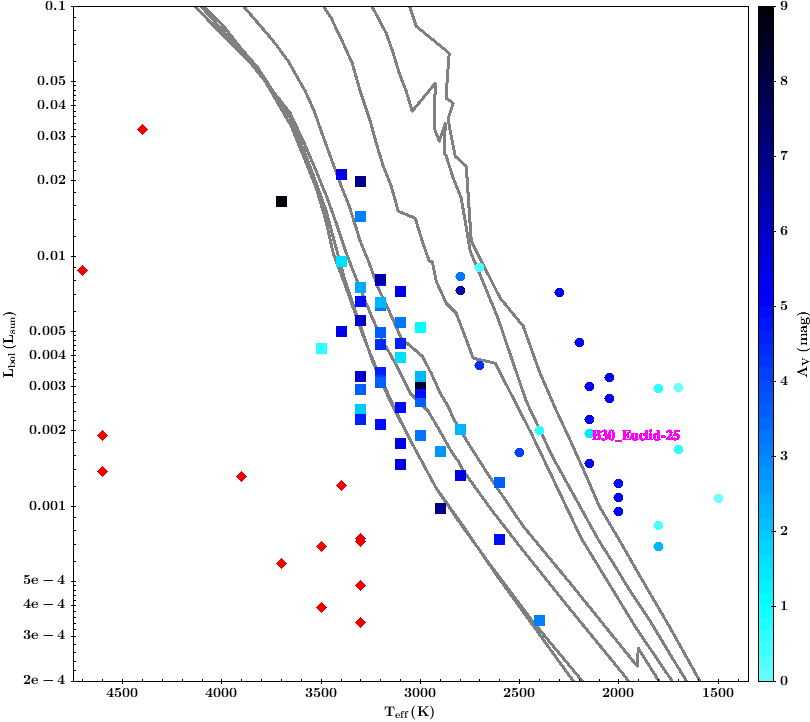}
      \caption{
        HRD for the candidate Barnard 30 candidate members based on our photometric selection.
        Rejected candidates, based on the derived T$_{\rm eff}$ and luminosities, appear as red solid diamond symbols.
        Possible candidate members are displayed as solid squares, whereas probable members, closer to the 3 Myr isochrones, appear as solid circles.
        The color scale indicates the derived reddening value A$_V$.
   The light gray lines correspond to a BT-settl isochrones with 1, 3, 10, 50, 100, 700, and 10,000 Myr \citep{Allard2013,Allard2014}.   
   }
         \label{F_HRD_Av}
   \end{figure}

We have used VOSA \citep[Virtual Observatory SED Analyzer;][]{Bayo2008.1, Bayo2017} to analyse the complete photometric data of the 94 Euclid candidates, ranging from 0.6 to 22 $\mu$m (WISE W4), when available. We have fitted the BT-Settl models from \citet{Allard2013,Allard2014}. We assumed a distance of 400 pc, a surface gravity value of logg=3.5, and a reddening range of A$_V$=0-10.0 mag. We rejected fits when the number of photometric data-points was less than eight measurements (11 objects).

Our results are shown in Figure~\ref{F_HRD_Av}, where we present the HR diagram for the candidates for membership of B30 dark cloud (94 in total, 83 with estimated values for $T_{\rm eff}$ and $L_{\rm bol}$) and highlight with solid red diamonds those whose membership has been rejected because they are too hot and/or too faint (16 in total). The figure also includes theoretical isochrones. We note that part of the remaining candidate members are not at the top of the 3 Myr isochrone, as our analysis is subject to significant uncertainties, such as individual distances (we have assumed 400\,pc for all of them since they are too faint to be detected by {\em Gaia}), reddening derived by VOSA (both from circumstellar discs or envelopes and from intra-cluster absorption) and photometric errors in some cases. Thus, we have redefined our membership classification for  the remaining candidates: 
 probable members for those close or above the 3\,Myr isochrone (23 objects), and possible members for those  between (and close to) the 50\,Myr and 10 Gyr isochrones at the dark cloud distance (44 objects).
For the former, many of the candidates lie above the 1\,Myr isochrone. An alternative possibility would be that they are at a very early evolutionary stage. 
For the latter, they show warmer temperatures and seem under-luminous relative to the expected location of the cluster sequence. Although their position in the HR diagram might indicate that most of them are background contaminants, we have kept them as possible members to unveil their true nature in future spectroscopic or astrometric surveys. Thus,  this comparison suggests that our pollution rate is higher than 15\% and 33\% for those objects coming from selection \#1 and \#2, respectively.

\begin{figure}
   \centering
\includegraphics[width=8.8cm]{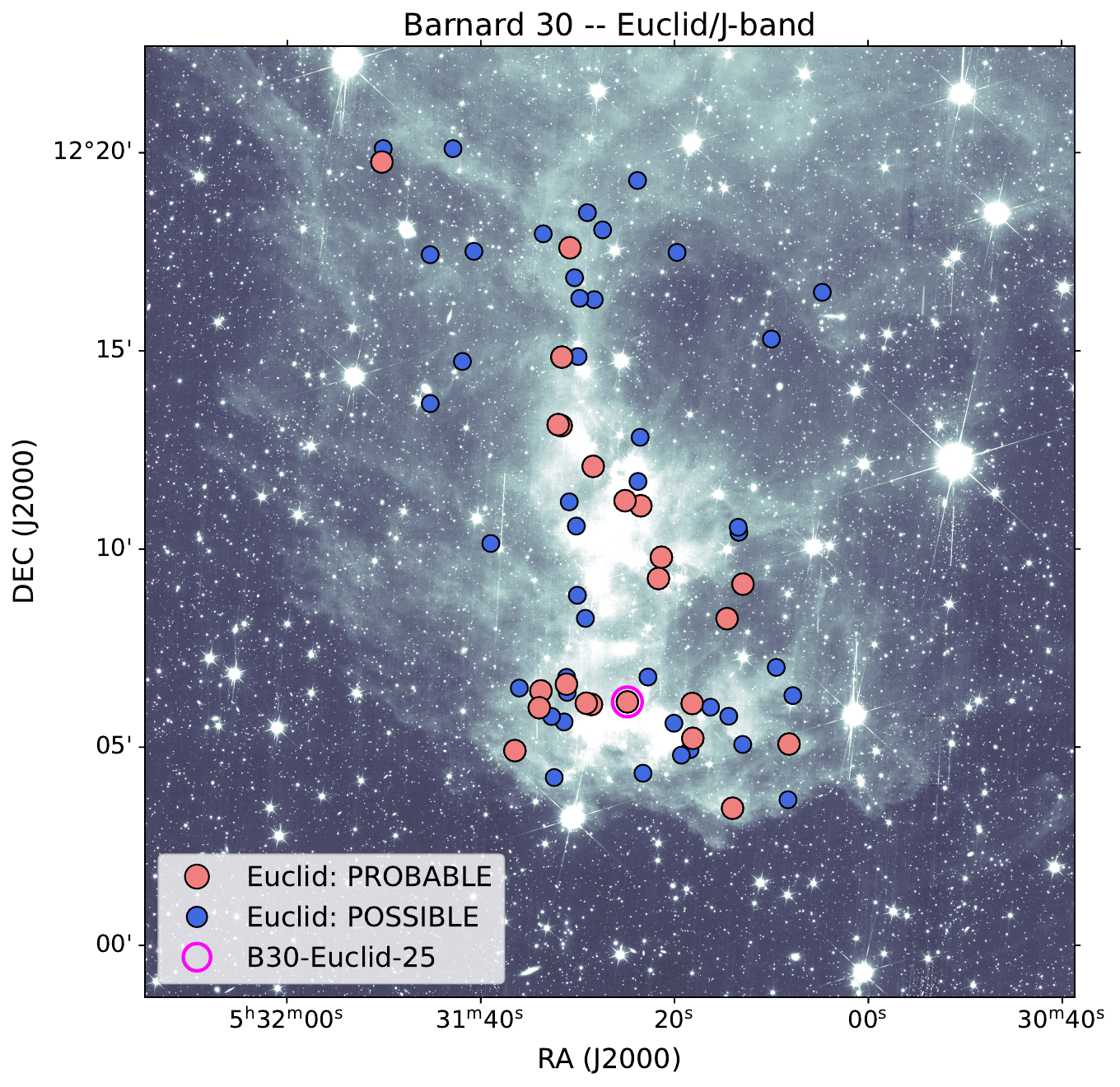}
   \caption{
Location of the 67 Euclid B30 candidate members selected after SED fitting. The sources are overlaid on an Euclid image taken with the \JE filter. Our candidate members are shown as coral (probable) and blue (possible) circles. The confirmed candidate B30-Euclid-25 is highlighted with a magenta circle.
   }

         \label{F_Image_Candidates}
   \end{figure}

In any case, we have been able to select a final sample of 67 candidate members of the Barnard 30 dark cloud and 23 look, indeed, as likely members. They are represented in Fig.~\ref{F_Image_Candidates}.
The results and the selection are listed in  Tab.~\ref{TAB_VOSA_HRD_Yes} --possible and probable members- and Tab.~\ref{TAB_VOSA_HRD_No} --rejected candidates.

   \begin{figure*}
   \centering
   \includegraphics[width=8.8cm]{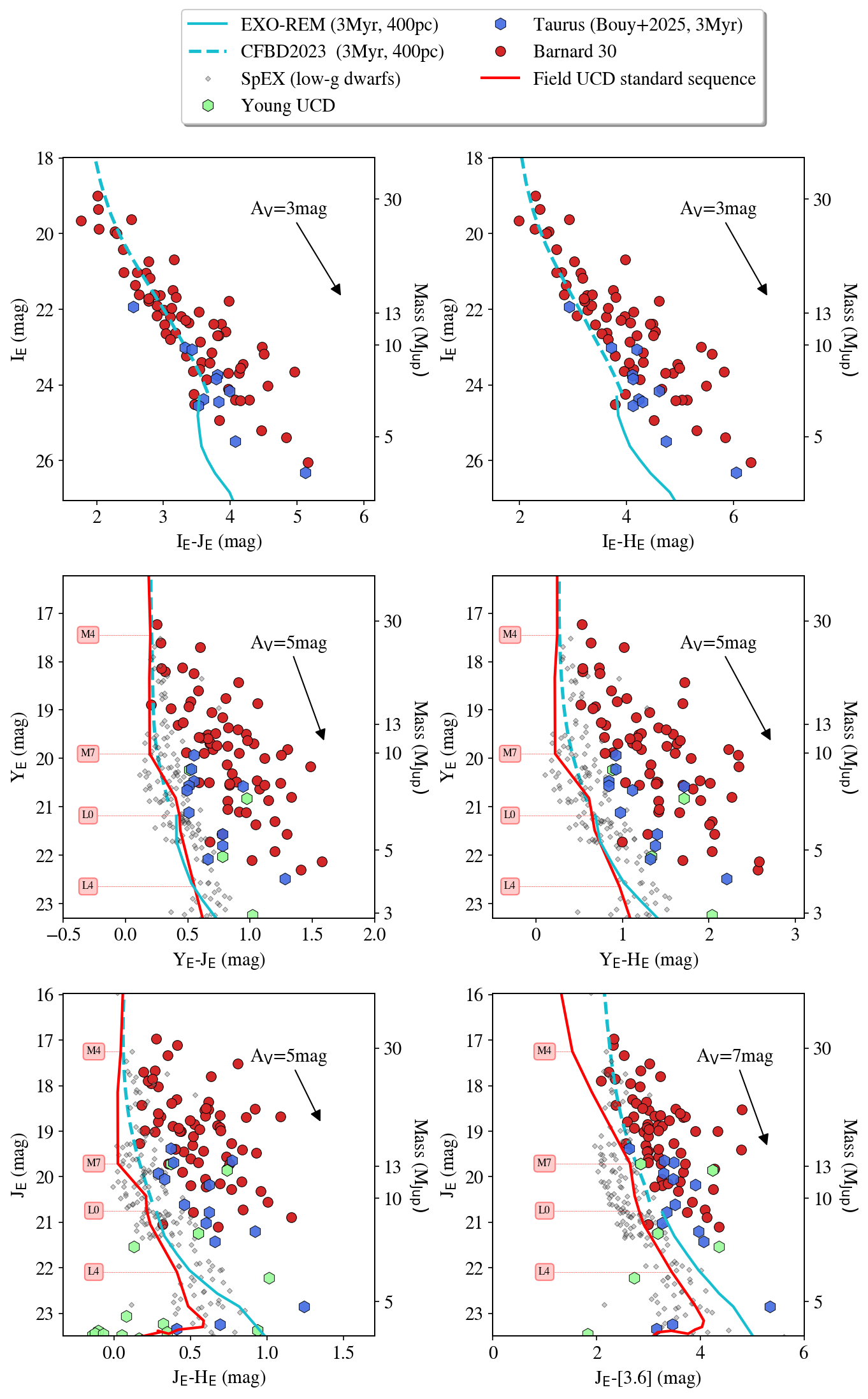}
   \caption{
     Color-Magnitude Diagrams of the 67 Euclid B30 candidate members  selected after SED fitting. For comparison, we include several young ultra-cool objects, and members of the Taurus region (\citealt{Bouy2025-EUCLID}) confirmed by proper motion . The field UCD standard sequence has been adopted from \citet{Bouy2025-EUCLID}.}
         \label{F_CM_Comparison}
   \end{figure*}

\subsection{Comparison with other young, cool populations}
\label{CMD}

We compared  our sample of 67 candidate members to similar substellar datasets, with the results displayed in Fig.~\ref{F_CM_Comparison}. In particular, we selected faint members from the following regions:
the Taurus LDN1495 dark cloud, shown as orange hexagons, and with available Euclid photometry \citep{Bouy2025-EUCLID};
a collection of known ultra-cool dwarfs classified as low or intermediate gravity in the IRTF/SpeX Ultracool Dwarf Library \citep{Burgasser2017-Spectra}, shown as grey diamonds; a compilation of young L and T-dwarfs from various sources \citep{Liu2013_PSO, Schneider2016_WISE224724, Miles2023_VHS1256, Luhman2023_TWA27, Zhang2021_Tdwafs}, shown as green hexagons. For these two last samples of cool dwarfs, their photometry in the Euclid bands was derived from their spectra as explained in \citet{Bouy2025-EUCLID}.

It is important to note that the Taurus 
sample was also observed as part of the \textit{Euclid} ERO program, and the membership criteria was based on proper motions. 
Therefore, the selection criteria for this dataset was independent of the \textit{Euclid} photometric properties used in our B30 analysis.

Despite the uncertainties on the precise location of the B30 sequence--due to variable and inhomogeneous extinction within the dark cloud--there is excellent agreement between the photometric positions of all these samples of young ultra-cool stellar and substellar objects. Figure~\ref{F_CM_Comparison} also provides an estimate of mass along the right-hand axis. If confirmed, and depending on the final extinction correction applied, our faintest candidate members would correspond to masses in the range of 4–5 M$_{\text{Jup}}$.

\subsection{B30-Euclid-25: Spectroscopic and photometric validation}
\label{SpectrocopicConfirmation}

We have been able to confirm the ultra-cool nature of one of our candidates, specifically B30-Euclid-025, which is included in the sample of 23 probable members. Our analysis includes both a comprehensive comparison of spectral energy distributions (SEDs) with models and spectral templates, as well as low-resolution spectroscopy in the near infrared collected with the Spanish 10-metre Gran Telescopio Canarias (GTC).

\subsubsection{Spectral Energy Distribution}
\label{SEDConfirmation}

  \begin{figure}
   \centering
   \includegraphics[width=8.8cm]{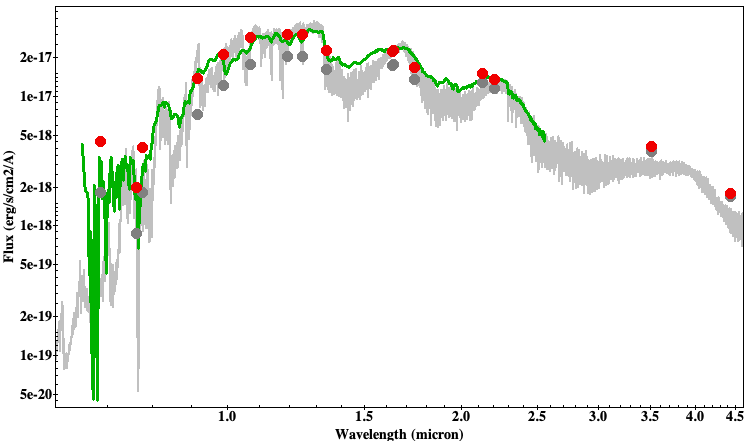}
      \caption{
   SED of B30-Euclid-025 and a comparison with a L1 spectral template  from the SPEX database
   (green line, \citealt{Burgasser2017-SPLAT}).
   Solid dark gray  and red circles represent the original and the unreddened  photometric data for our B30 candidate member, respectively (A$_V$=1.375 mag).    The light gray line corresponds to a BT-settl model with 2200 K and logg=3.5  \citep{Allard2013,Allard2014}.   
   }
         \label{F_SEDCompS}
   \end{figure}

We have used VOSA \citep{Bayo2008.1, Bayo2017} to perform a more detailed analysis of B30-Euclid-25, covering wavelengths from 0.6 to 4.5 $\mu$m. We have fitted the BT-Settl models from \citet{Allard2013,Allard2014}. Our initial parameter space includes d=400 pc, $T_{\rm eff}$= 4000-2000\,K, log\,g=2.4-5.5, and a reddening A$_V$=0-6.0 mag. The best fit is obtained with a single object with $T_{\rm eff}$=2200 K, log\,g=3.5, A$_V$=1.375 mag, $L_{\rm bol}$=1.942$\times$10$^{-3}$ $L_\odot$ (Fig.~\ref{F_SEDCompS}). There is an excess at the mid-infrared beyond 3.5 $\mu$m. This result places B30-Euclid-25 as a super-luminous UCD compared to the association isochrone at $\sim$3 Myr. 
However, the fit would be similar with two equal-mass binary (same temperature and half the luminosity). 
We have also fitted a spectral template from \citet{Burgasser2017-SPLAT}  with VOSA. The best result indicates a spectral type of L1 (shown as a green line in the diagram), cooler than the value expected from the T$_{\rm eff}$ derived from the SED fit.

As an alternative, we have forced a binary fit with logg=3.5 (the expected value for the age of the dark cloud B30) and A$_V$=0.0 (to minimise the number of free parameters). The fit, shown in Figure~\ref{F_SEDCompBin}, is of high quality and produces $L_{\rm bol}$=1.052$\times$10$^{-3}$ $L_\odot$  and $L_{\rm bol}$=5.713$\times$10$^{-4}$ $L_\odot$, respectively.  These values, subject to the relevant uncertainties, indicate that the hotter component ($T_{\rm eff}$ = 2200\,K)  would be a low-mass brown dwarf, while the cooler one ($T_{\rm eff}$ = 1700\,K) would be a planetary-mass object.

\subsubsection{Spectroscopic confirmation and characterization}
\label{SpectrocopicConfirmation}

   \begin{figure}
   \centering
   \includegraphics[width=8.8cm]{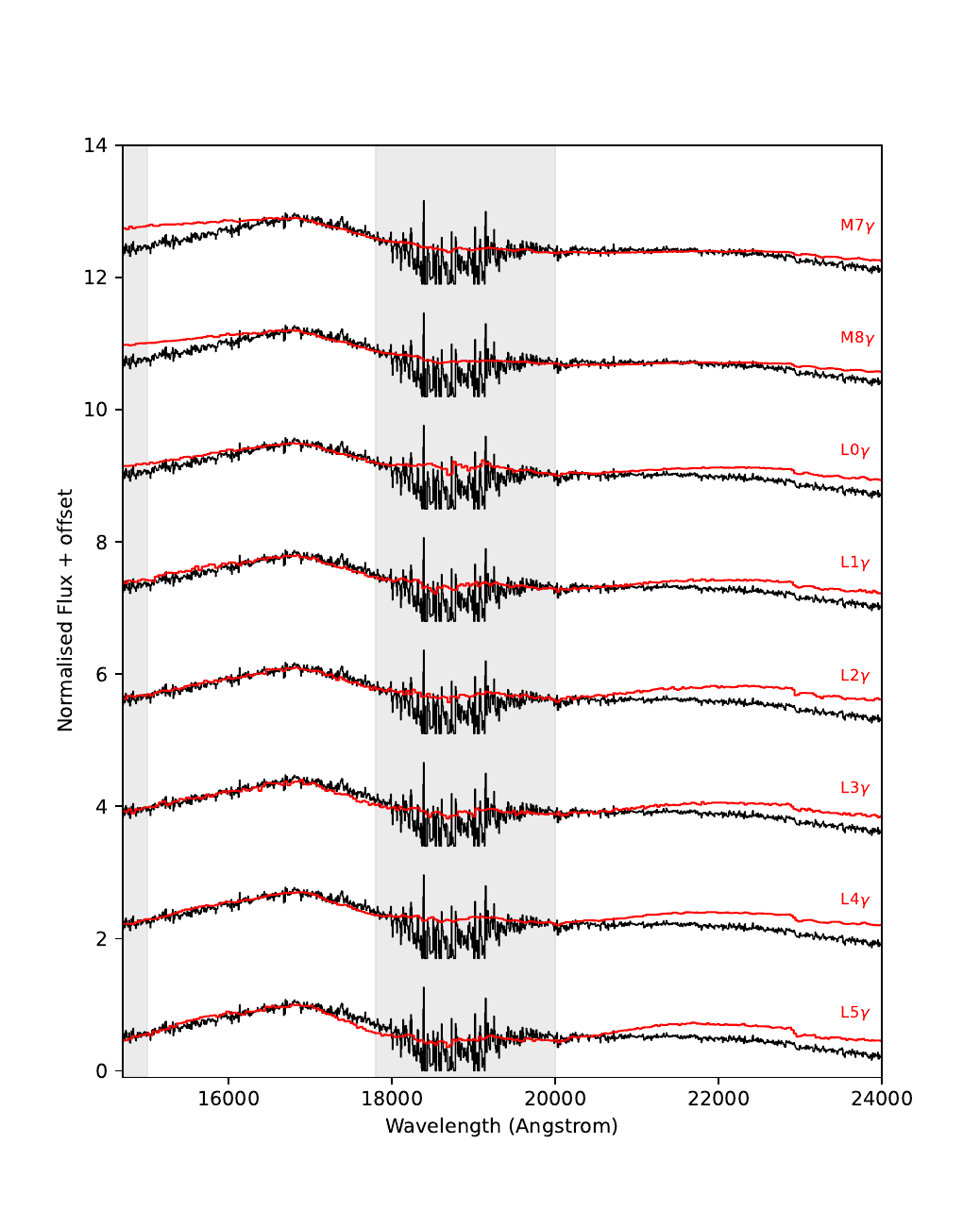}
      \caption{
   The GTC/EMIR spectrum of B30-Euclid-025 and a comparison with a set of low-gravity spectral templates extracted from the SPEX database
    (\citealt{Burgasser2017-SPLAT}).
   }
         \label{F_SpectraCompLG}
   \end{figure}

   \begin{figure}
   \centering
   \includegraphics[width=8.8cm]{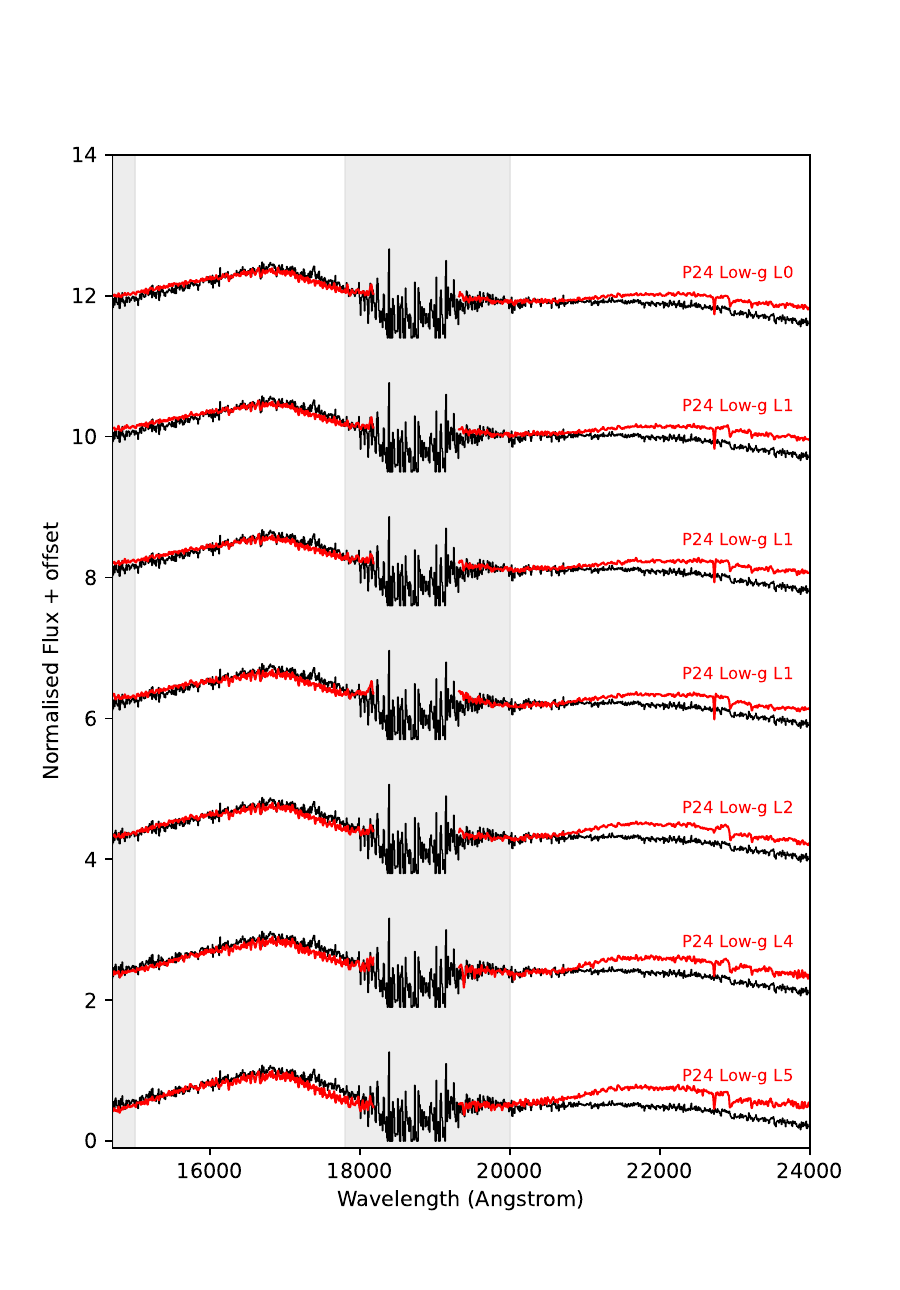}
      \caption{
   The GTC/EMIR spectrum of B30-Euclid-025 and a comparison with a set of young spectral templates extracted from \citet{Piscarreta2024-LT}.
   }
         \label{F_SpectraComp_Piscarreta}
   \end{figure}
   
The NIR spectrum of B30-Euclid-025, obtained using EMIR on GTC, was compared against several spectral templates spanning a range of spectral types and surface gravities. 
These templates include low-resolution, low-gravity and Main Sequence spectra from the SPLAT database \citep{Burgasser2017-SPLAT}, and young (ages $<$ 50\,Myr) spectral templates from \citet{Piscarreta2024-LT}.
The comparison was performed after dereddening the observed spectrum using the extinction law from \citet{Cardelli1989} and considering four extinction values: $A_V = 0.0$, $3.0$, $5.0$, and $10.0$ mag. 

Figure~\ref{F_SpectraCompLG} shows the comparison with low-gravity templates from \citet{Burgasser2017-SPLAT}. The best spectral match was obtained for $A_V = 3.0$ mag, and a spectral type of approximately L2. We then refined the fit including a finer grid of $A_V$ values ranging between 1.5 and 4.5 mag with a  step of 0.5 mag,  obtaining a best fit for $A_V$ values of 2.5-3.0\,mag and a spectral type of L2-L3.
We note however that, while the fit in the $H$ band is relatively robust, discrepancies are noticeable in the $K$ band. 

The same occurs when comparing the NIR spectrum with young templates from \citet{Piscarreta2024-LT}. As seen in
Figure~\ref{F_SpectraComp_Piscarreta}, the best match is obtained for L1- L2 objects, with a clear mismatch in the K-band region.
As a result of these comparisons, we can conclude that the H-band spectrum of our target is consistent with an early L dwarf with low gravity, while the K-band is more consistent with a late-M. Our tentative classification is L1-L3 with a moderate reddening of A$_V$ $\sim$3.0\,mag.

We have derived some spectral indexes that are commonly used as spectral type and youth indicators. The H$_2$O index proposed in \citet{Allers2007} is consistent with the estimated L2 spetral type (value between 12-13), while the $H_{\rm cont}$ index from \citet{Allers2013} is too low to be consistent with a low-gravity object (value of $\sim$0.78). We have also derived the TLI-g index proposed by \citet{Almendros2022}, displaying a value of $\sim$0.83, which seems more consistent with low-gravity objects  \citep[see e.g.,][]{Bouy2022-UpSco}.

Although low-resolution spectroscopy is insufficient to definitively confirm the membership of B30-Euclid-025 in the B30 region, several pieces of evidence would support its association. These include the characteristic triangular shape in the $H$-band spectrum, the resemblance to low-gravity L dwarfs, and consistent photometric properties. Collectively, these features suggest that the object is likely a $\sim$3 Myr-old  brown dwarf physically associated with the B30 dark cloud. If this classification is confirmed, the estimated mass would fall within the range of $15$--$20~M_{\mathrm{Jup}}$.

\subsection{Spatial distribution}
\label{SpatianDistribution}

To conclude our analysis, we present the full list of candidate members overlaid on a \textit{Euclid} $J$-band image. The spatial distribution of these objects is shown in Figure~\ref{F_Image_Candidates}. It is important to note that our selection was intentionally restricted to the interior of the dark cloud to minimize contamination, particularly from distant extra-galactic sources.

The distribution of our candidates closely follows the morphology of the dark cloud. However, a few sources are located in regions of only moderate extinction. These objects are especially promising for further study and should be prioritized for spectroscopic follow-up.

\section{Conclusions}

We have presented \textit{Euclid} Early Release Observations of the B30 dark cloud in the Lambda Orionis star-forming region. Our main conclusions can be summarized as follows:

\begin{enumerate}
\item We have identified a large sample of 23 probable very low-mass members of the B30 dark cloud.
\item Even after accounting for reddening and contamination by spurious sources—primarily unresolved background galaxies—our survey appears to reach down to a few Jupiter masses, approximately 4–5 M$_{\text{Jup}}$, potentially approaching the cut-off of the initial mass function (IMF).
\item Additional deep imaging, particularly in filters such as $i$, $z$, and $Ks$, is necessary to validate some of the faintest candidates.
\item We have obtained a low-resolution near-IR spectrum of one target. Our comparison with different ultra-cool dwarf templates suggests it is a L2 dwarf with low gravity and moderate extinction. This initial result validates our photometric selection methodology.
\item We have identified a handful of wide binary candidates among the low-mass population.
\item Our sample includes several ideal targets for spectroscopic follow-up, particularly with high-sensitivity facilities such as the James Webb Space Telescope (JWST).
\item The confirmation of even a subset of these candidates would place strong constraints on the formation mechanisms of substellar and planetary-mass objects in young star-forming regions.
\end{enumerate}

   \begin{acknowledgements}
     This work has made use of the Early Release Observations
(ERO) data from the \textit{Euclid} mission of the European Space Agency (ESA),
     2024, \textit{Euclid}, available at
     https://euclid.esac.esa.int/dr/ero/.  We thank I. Baraffe and M. Phillips for making available a digitised version of their isochrones in the \textit{Euclid} passbands.
DB and NH are  funded by grants by grants No. PID2019-107061GB-C61 and PID2023-150468NB-I00 by the Spain Ministry of Science and Innovation/State Agency of Research MCIN/AEI/ 10.13039/501100011033.
ELM, MŽ, CDT and JZ  are supported by the European Research Council Advanced grant SUBSTELLAR, project number 101054354.
  This research has made use of the NASA/IPAC
Infrared Science Archive, which is funded by the National Aeronautics and
Space Administration and operated by the California Institute of Technology.
This research has made use of the Spanish Virtual Observatory (https://svo.
cab.inta-csic.es) project funded by MCIN/AEI/10.13039/501100011033/
through grant PID2020-112949GB-I00. This work has benefited from The UltracoolSheet
at http://bit.ly/UltracoolSheet, maintained by Will Best,
Trent Dupuy, Michael Liu, Rob Siverd, and Zhoujian Zhang. Based in part on
data collected at Subaru Telescope which is operated by the National Astronomical Observatory of Japan and obtained from the SMOKA, which is operated
by the Astronomy Data Center, National Astronomical Observatory of Japan.
This research has made use of the VizieR catalogue access tool, CDS, Strasbourg,
France. The original description of the VizieR service was published in
A\&AS 143, 23. This work made use of Topcat (\citealt{Taylor2005_Topcat}).
\end{acknowledgements}

%
%


\small
%
\bibliographystyle{aa} 
\bibliography{Barrado_B30_Euclid_bibliography}



%
%

\clearpage

\setcounter{table}{0}
%
%
\input{0_Tables/T01_B30_VOSA_bestfitp_mod_Yes}




\appendix

\section{Additional  Color-Magnitude diagrams}

      \begin{figure*}
   \centering
\includegraphics[width=8.8cm]{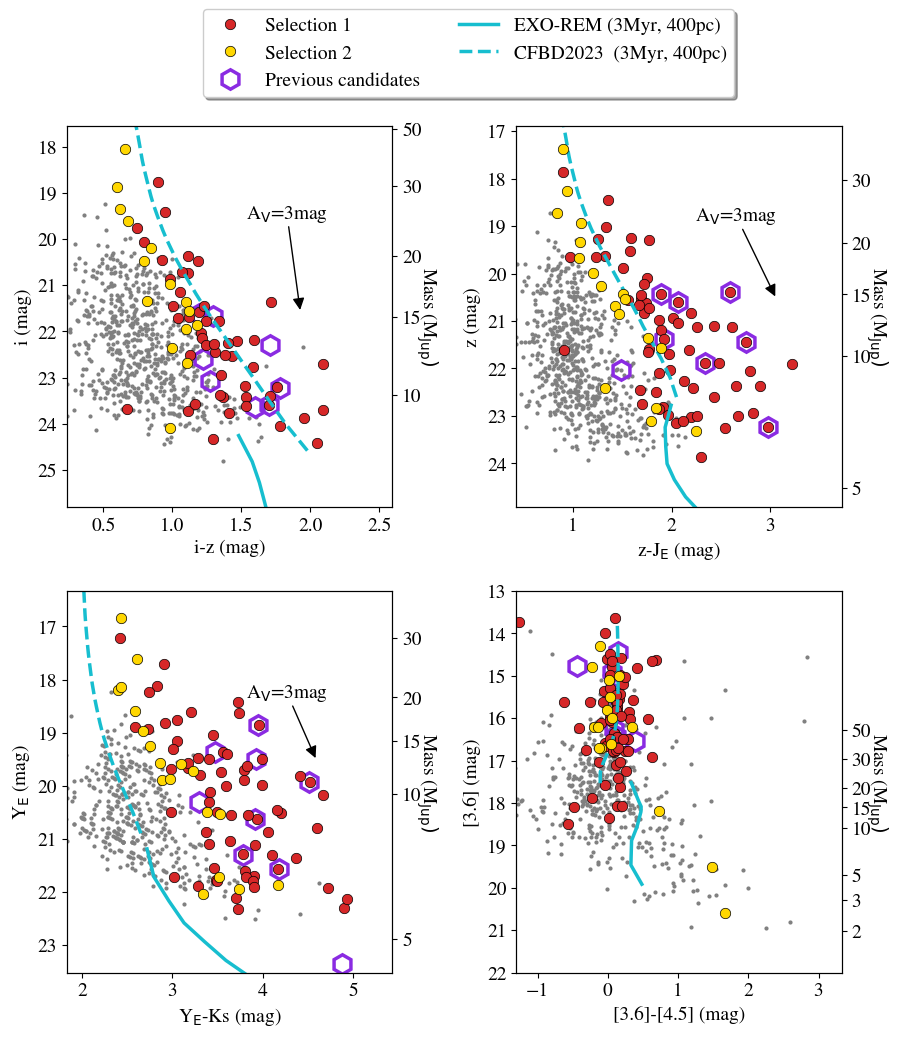} 
   \caption{
     Optical/near-IR color-magnitude diagrams. The initial B30 sample is shown as gray dots, while our candidate members are indicated by red (selection \#1) and orange (selection \#2) circles. Previously known B30 candidates from \citet{Huelamo2017-ALMA-B30} and \citet{Barrado2018_B30} are marked with purple hexagons. A 3\,Myr isochrone at 400\,pc is plotted as a light blue curve, dashed for CFBD2023 and solid for EXO-REM models, respectively. The corresponding mass scale is shown on the right-hand axis. Extinction vectors are also included.
   }
         \label{F_CM_OptIR}
   \end{figure*}

   \begin{figure}
   \centering
   \includegraphics[width=8.8cm]{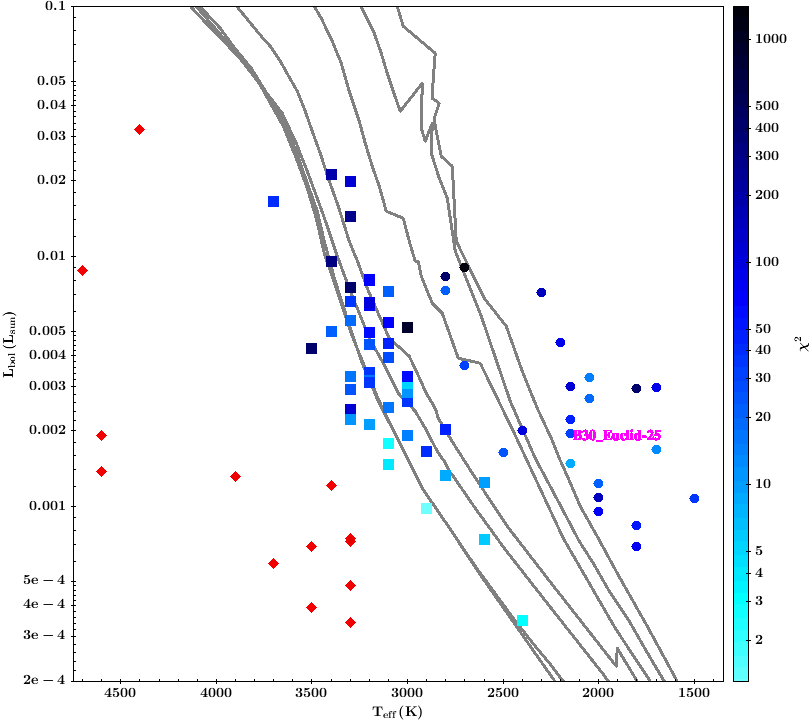}
      \caption{
        HRD for the candidate Barnard 30 candidate members based on our photometric selection.
        Rejected candidates, based on the derived Teff and luminosities, appear as red solid diamond symbols.
        Possible candidate members are displayed as solid squares, whereas probable members, closer to the 3 Myr isochornes, appear as solid circles.
        The color scale indicates the derived $\chi^2$.
   The light gray lines correspond to a BT-settl isochrones with 1, 3, 10, 50, 100, 700, and 10,000 Myr \citep{Allard2013,Allard2014}.   
   }
         \label{F_HRD_Chi2}
   \end{figure}

   \begin{figure}
   \centering
   \includegraphics[width=8.8cm]{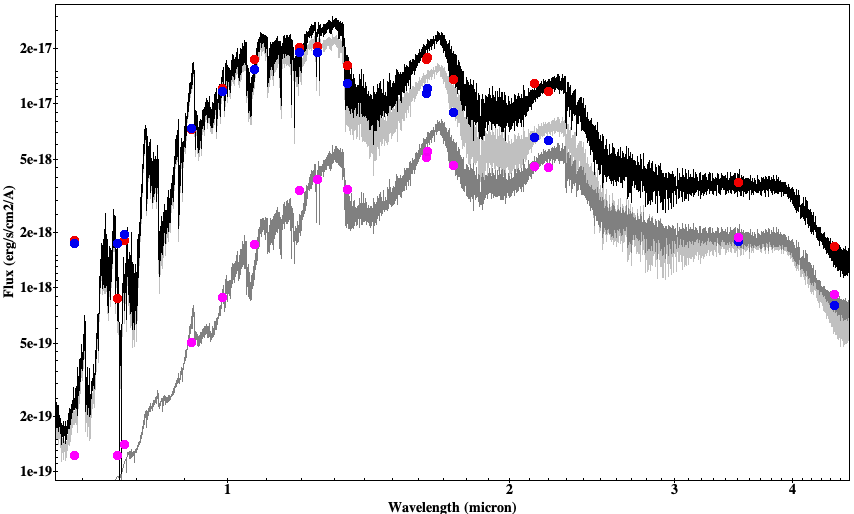}
      \caption{
   Spectral Energy Distribution for B30-Euclid-025, assuming it is composed by two UCDs with logg=3.5 and A$_V$=0.0 mag, and located at 400 pc.
   Solid magenta circles represent the original photometric data for our B30 candidate member, whereas blue and red circles correspond to the primary (T$_{eff}$=2200 K)
   and secondary  (T$_{eff}$=1700 K), respectively. Theoretical spectra from  \citep{Allard2013,Allard2014} for those temperatures and gravities are also included as light and dark gray, respectively (resolution=5000).
   }
         \label{F_SEDCompBin}
   \end{figure}

   \begin{figure}
   \centering
   \includegraphics[width=8.8cm]{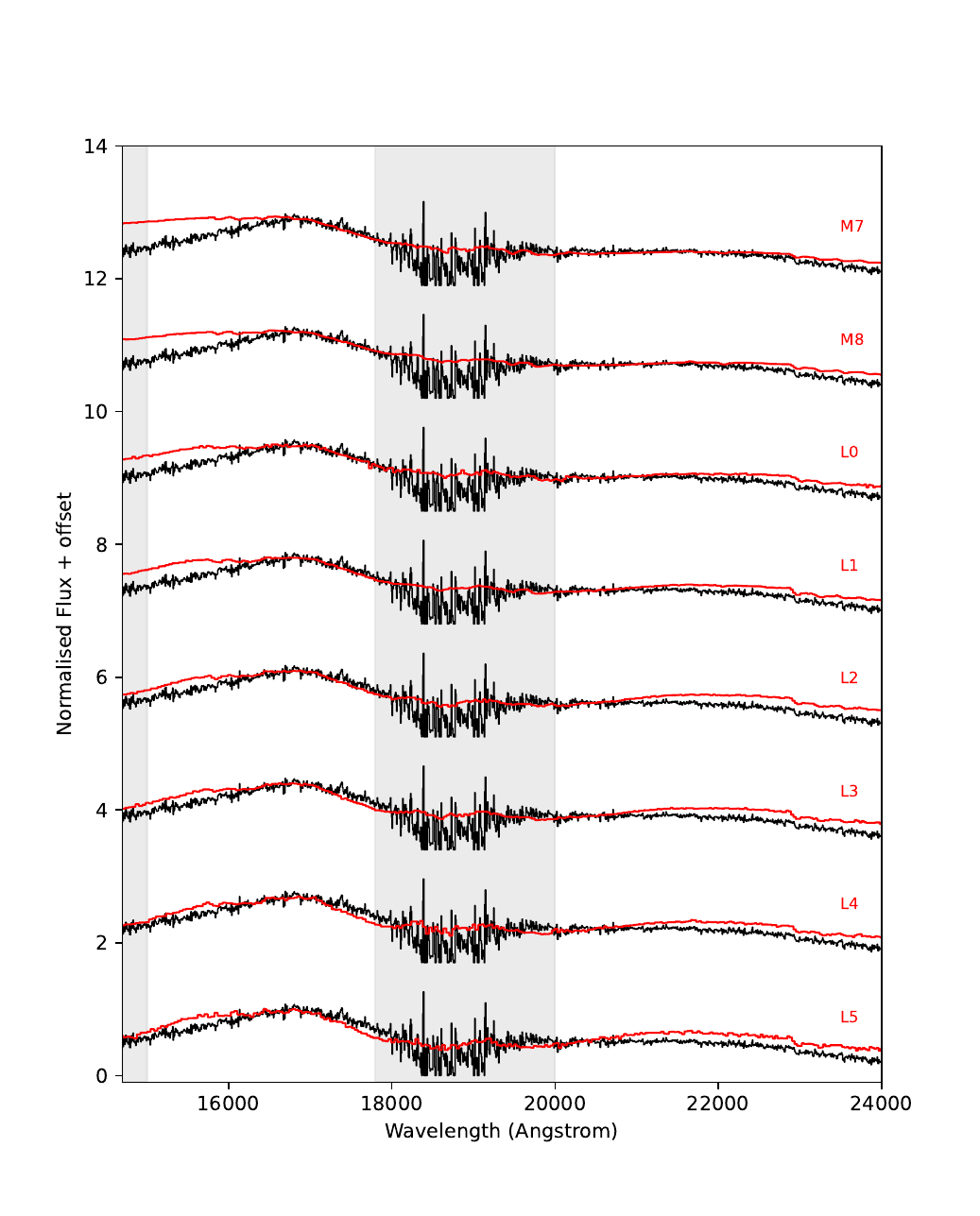}
      \caption{
   The GTC/EMIR spectrum of B30-Euclid-025 and a comparison with a set of Main Sequence spectral templates extracted from the SPEX database
    (\citealt{Burgasser2017-SPLAT}).
   }
         \label{F_SpectraComp_MS}
   \end{figure}

%
%


\setcounter{table}{0}
%
%
\input{0_Tables/T_A01_B30_HB_DR3_core.izYHKsSpitzerPM_sel_HB}


\setcounter{table}{1}
%
%
\input{0_Tables/T_A02_B30_HB_DR3_core.izYHKsSpitzerPM_mem_Final}


\setcounter{table}{2}
%
%
\input{0_Tables/T_A03_B30_VOSA_bestfitp_mod_No}


\end{document}

%% file: 0_Tables/T01_B30_VOSA_bestfitp_mod_Yes.tex
\tiny
\begin{table*} 
\tiny
\caption{Properties of Barnard 30  possible and probable  candidate members based on the analysis of the Spectral Energy Distribution.   }
\begin{tabular}{ l r r r r r r r r r l r } 
\hline
  \multicolumn{1}{c}{Object} &
  \multicolumn{1}{c}{RA} &
  \multicolumn{1}{c}{DEC} &
  \multicolumn{1}{c}{Teff} &
  \multicolumn{1}{c}{e\_Teff} &
  \multicolumn{1}{c}{Lbol} &
  \multicolumn{1}{c}{Lberr} &
  \multicolumn{1}{c}{Av} &
  \multicolumn{1}{c}{$\chi^2$} &
  \multicolumn{1}{c}{Fobs/Ftot} &
  \multicolumn{1}{c}{Mem?} &
  \multicolumn{1}{c}{Sp.Type} \\ 
%
  \multicolumn{1}{c}{} &
  \multicolumn{1}{c}{(deg)} &
  \multicolumn{1}{c}{(deg)} &
  \multicolumn{1}{c}{(K)} &
  \multicolumn{1}{c}{(K)} &
  \multicolumn{1}{c}{(L$_\odot$)} &
  \multicolumn{1}{c}{(L$_\odot$)} &
  \multicolumn{1}{c}{(mag)} &
  \multicolumn{1}{c}{} &
  \multicolumn{1}{c}{} &
  \multicolumn{1}{c}{} &
  \multicolumn{1}{c}{(template)} \\ 
\hline                                                                                                                                                                                                 
   B30\_Euclid-01 & 82.7841   & 12.084713 & 2050 & 25 & 0.002697 & 8.608703E-6 & 5.225  & 19.2    & 0.778  & Prob & Opt:M8V         \\  
   B30\_Euclid-02 & 82.78457  & 12.061252 & 3100 & 50 & 0.005412 & 8.521616E-6 & 3.3    & 70.3    & 0.833  & Poss & Opt:M5V         \\  
   B30\_Euclid-03 & 82.78958  & 12.116933 & 3100 & 50 & 0.007202 & 8.442275E-6 & 5.5    & 22.6    & 0.827  & Poss & Opt:M4V         \\  
   B30\_Euclid-04 & 82.80389  & 12.151958 & 2800 & 50 & 0.008264 & 9.029737E-6 & 3.3    & 410.3   & 0.823  & Prob & Opt:M5V         \\  
   B30\_Euclid-05 & 82.80403  & 12.084534 & 3200 & 50 & 0.008002 & 8.557211E-6 & 4.95   & 75.2    & 0.831  & Poss & Opt:M4V         \\  
   B30\_Euclid-06 & 82.80564  & 12.173745 & 3000 & 50 & 0.002619 & 7.784802E-6 & 3.575  & 34.4    & 0.83   & Poss & Opt:M5V         \\  
   B30\_Euclid-07 & 82.805916 & 12.17586  & 2600 & 50 & 7.37E-4  & 6.768503E-6 & 4.95   & 5.6     & 0.753  & Poss & Opt:M5V         \\  
   B30\_Euclid-08 & 82.808395 & 12.057712 & 2300 & 25 & 0.007136 & 9.933619E-6 & 5.225  & 155.7   & 0.836  & Prob & Opt:M6V         \\  
   B30\_Euclid-09 & 82.81002  & 12.096402 & 3300 & 50 & 0.019936 & 7.235967E-5 & 7.0    & 120.2   & 0.866  & Poss & Opt:M5V         \\  
   B30\_Euclid-10 & 82.810745 & 12.137424 & 2400 & 37 & 0.002009 & 9.049523E-6 & 0.55   & 93.4    & 0.915  & Prob & Opt:L1          \\  
   B30\_Euclid-11 & 82.825485 & 12.087104 & 2800 & 50 & 0.007289 & 8.0513655E-6& 6.5    & 21.7    & 0.824  & Prob & Opt:M5V         \\  
   B30\_Euclid-12 & 82.82573  & 12.101757 & 1800 & 25 & 8.39E-4  & 6.7715537E-6& 0.275  & 51.5    & 0.614  & Prob & Opt:L8,NIR:L8   \\  
   B30\_Euclid-13 & 82.826645 & 12.082364 & 3000 & 50 & 0.002854 & 6.9452535E-6& 5.0    & 11.9    & 0.83   & Poss & Opt:M5V         \\  
   B30\_Euclid-14 & 82.8304   & 12.080015 & 3300 & 50 & 0.014347 & 9.240884E-6 & 3.025  & 276.0   & 0.819  & Poss & Opt:M4V         \\  
   B30\_Euclid-15 & 82.8336   & 12.093461 & 3200 & 50 & 0.003418 & 7.887208E-6 & 4.675  & 35.4    & 0.879  & Poss & Opt:M5V         \\  
   B30\_Euclid-17 & 82.83898  & 12.163276 & 2150 & 25 & 0.002997 & 8.114288E-6 & 5.225  & 114.5   & 0.794  & Prob & Opt:L4          \\  
   B30\_Euclid-19 & 82.84024  & 12.154336 & 2050 & 25 & 0.003278 & 8.55449E-6  & 5.225  & 14.5    & 0.78   & Prob & Opt:L4          \\  
   B30\_Euclid-20 & 82.84477  & 12.112886 & 3100 & 50 & 0.004472 & 8.473943E-6 & 4.675  & 44.4    & 0.886  & Poss & Opt:M5V         \\  
   B30\_Euclid-21 & 82.84691  & 12.072429 & 3200 & 50 & 0.004946 & 8.330678E-6 & 3.575  & 57.2    & 0.825  & Poss & Opt:M4V         \\  
   B30\_Euclid-22 & 82.84782  & 12.184915 & 2200 & 25 & 0.00453  & 8.002203E-6 & 5.225  & 90.6    & 0.809  & Prob & Opt:M6V         \\  
   B30\_Euclid-23 & 82.849045 & 12.195134 & 3400 & 50 & 0.004986 & 8.59569E-6  & 5.5    & 22.4    & 0.859  & Poss & Opt:M5V         \\  
   B30\_Euclid-24 & 82.84922  & 12.321717 & 3300 & 50 & 0.007501 & 6.299313E-5 & 2.475  & 393.8   & 0.935  & Poss & Opt:M5V         \\  
   B30\_Euclid-25 & 82.85361  & 12.102432 & 2150 & 25 & 0.001952 & 7.650488E-6 & 1.375  & 30.1    & 0.805  & Prob & Opt:L4          \\  
   B30\_Euclid-27 & 82.8546   & 12.187048 & 1800 & 25 & 0.002945 & 1.3122318E-5& 0.55   & 400.2   & 0.888  & Prob & Opt:L6          \\  
   B30\_Euclid-28 & 82.86426  & 12.300935 & 2600 & 50 & 0.001244 & 7.859989E-6 & 3.575  & 9.6     & 0.833  & Poss & Opt:M6V         \\  
   B30\_Euclid-30 & 82.86786  & 12.271597 & 3100 & 50 & 0.00178  & 7.887774E-6 & 5.5    & 2.9     & 0.833  & Poss & Opt:M5V         \\  
   B30\_Euclid-31 & 82.868324 & 12.201486 & 1800 & 25 & 6.93E-4  & 7.729256E-6 & 2.2    & 78.3    & 0.735  & Prob & Opt:L6          \\  
   B30\_Euclid-33 & 82.86908  & 12.101358 & 1700 & 25 & 0.001682 & 7.849315E-6 & 0.55   & 12.2    & 0.513  & Prob & Opt:L6          \\  
   B30\_Euclid-35 & 82.87126  & 12.10188  & 1700 & 25 & 0.002974 & 8.8238285E-6& 0.0    & 92.5    & 0.528  & Prob & Opt:T2,NIR:T2   \\  
   B30\_Euclid-36 & 82.87167  & 12.137554 & 3300 & 50 & 0.006572 & 8.302692E-6 & 4.95   & 37.0    & 0.822  & Poss & Opt:M5V         \\  
   B30\_Euclid-38 & 82.8741   & 12.272169 & 3300 & 50 & 0.003314 & 8.461092E-6 & 6.0    & 16.3    & 0.872  & Poss & Opt:M4V         \\  
   B30\_Euclid-39 & 82.8748   & 12.247682 & 3100 & 50 & 0.002474 & 8.198087E-6 & 5.0    & 16.4    & 0.883  & Poss & Opt:M5V         \\  
   B30\_Euclid-40 & 82.875114 & 12.147262 & 3200 & 50 & 0.00319  & 7.924496E-6 & 4.125  & 17.2    & 0.822  & Poss & Opt:M4V         \\  
   B30\_Euclid-41 & 82.875565 & 12.176378 & 3000 & 50 & 0.003084 & 6.875764E-6 & 8.5    & 4.9     & 0.759  & Poss & Opt:M5V         \\  
   B30\_Euclid-43 & 82.876335 & 12.280811 & 2800 & 50 & 0.00133  & 6.2427475E-6& 5.5    & 8.7     & 0.807  & Poss & Opt:M5V         \\  
   B30\_Euclid-44 & 82.877495 & 12.292957 & 3700 & 50 & 0.016493 & 1.3050089E-5& 9.0    & 42.9    & 0.811  & Poss & Opt:M4V         \\  
   B30\_Euclid-45 & 82.8783   & 12.293422 & 2150 & 25 & 0.002226 & 8.434887E-6 & 5.225  & 49.5    & 0.801  & Prob & Opt:L4          \\  
   B30\_Euclid-46 & 82.878624 & 12.186555 & 3100 & 50 & 0.001474 & 7.500576E-6 & 5.5    & 3.7     & 0.868  & Poss & Opt:M5V         \\  
   B30\_Euclid-48 & 82.879745 & 12.11276  & 3300 & 50 & 0.005528 & 8.154679E-6 & 6.0    & 18.2    & 0.824  & Poss & Opt:M4V         \\  
   B30\_Euclid-49 & 82.87982  & 12.109877 & 1500 & 37 & 0.001076 & 6.818044E-6 & 0.0    & 36.4    & 0.367  & Prob & Opt:L6          \\  
   B30\_Euclid-50 & 82.88083  & 12.094015 & 3200 & 50 & 0.004451 & 8.367642E-6 & 4.4    & 25.5    & 0.827  & Poss & Opt:M4V         \\  
   B30\_Euclid-52 & 82.88178  & 12.247449 & 2700 & 50 & 0.003644 & 7.720408E-6 & 4.4    & 33.2    & 0.823  & Prob & Opt:M7V         \\  
   B30\_Euclid-53 & 82.88208  & 12.218594 & 2000 & 25 & 9.49E-4  & 6.8119234E-6& 5.225  & 74.6    & 0.836  & Prob & Opt:L6          \\  
   B30\_Euclid-56 & 82.88337  & 12.219047 & 2000 & 25 & 0.001084 & 7.1324284E-6& 5.225  & 141.4   & 0.809  & Prob & Opt:L6          \\  
   B30\_Euclid-58 & 82.8862   & 12.0963   & 3200 & 50 & 0.00805  & 9.458529E-6 & 6.5    & 72.7    & 0.885  & Poss & Opt:M5V         \\  
   B30\_Euclid-59 & 82.88982  & 12.2993   & 3400 & 50 & 0.021112 & 6.756197E-5 & 5.5    & 199.3   & 0.818  & Poss & Opt:M4V         \\  
   B30\_Euclid-60 & 82.8908   & 12.107047 & 2000 & 25 & 0.001231 & 7.0004226E-6& 4.95   & 21.2    & 0.739  & Prob & Opt:L8,NIR:T0   \\  
   B30\_Euclid-61 & 82.891495 & 12.099991 & 2150 & 25 & 0.001484 & 6.892342E-6 & 5.225  & 9.0     & 0.807  & Prob & Opt:L4          \\  
   B30\_Euclid-64 & 82.900085 & 12.108244 & 2900 & 50 & 9.77E-4  & 6.0553393E-6& 7.0    & 1.3     & 0.769  & Poss & Opt:M4V         \\  
   B30\_Euclid-65 & 82.902    & 12.081943 & 2500 & 50 & 0.001633 & 9.19779E-6  & 4.125  & 23.5    & 0.835  & Prob & Opt:M8V         \\  
   B30\_Euclid-66 & 82.91236  & 12.169068 & 2400 & 37 & 3.49E-4  & 4.8859038E-6& 3.025  & 3.1     & 0.75   & Poss & Opt:M7V         \\  
   B30\_Euclid-67 & 82.91965  & 12.291908 & 3200 & 50 & 0.006338 & 9.272295E-6 & 3.3    & 81.2    & 0.83   & Poss & Opt:M4V         \\  
   B30\_Euclid-68 & 82.92458  & 12.24562  & 3000 & 50 & 0.003299 & 8.335566E-6 & 2.2    & 63.2    & 0.896  & Poss & Opt:M5V         \\  
   B30\_Euclid-69 & 82.92866  & 12.335093 & 3200 & 50 & 0.006542 & 6.812318E-5 & 2.2    & 106.5   & 0.899  & Poss & Opt:M5V         \\  
   B30\_Euclid-71 & 82.95926  & 12.32948  & 2700 & 50 & 0.008965 & 6.896356E-5 & 0.275  & 1405.4 5& 0.935  & Prob & Opt:M6V         \\  
   B30\_Euclid-73 & 82.769714 & 12.274682 & 3300 & 50 & 0.002436 & 7.710472E-6 & 1.925  & 108.2   & 0.939  & Poss & Opt:M5V         \\  
   B30\_Euclid-74 & 82.782455 & 12.105072 & 3000 & 50 & 0.001909 & 8.328612E-6 & 3.3    & 14.4    & 0.832  & Poss & Opt:M5V         \\  
   B30\_Euclid-75 & 82.79162  & 12.255011 & 3500 & 50 & 0.004283 & 1.6863023E-5& 0.55   & 427.9   & 0.961  & Poss & Opt:M5V         \\  
   B30\_Euclid-79 & 82.81785  & 12.100211 & 3300 & 50 & 0.002212 & 7.703496E-6 & 4.675  & 11.7    & 0.827  & Poss & Opt:M4V         \\  
   B30\_Euclid-82 & 82.83222  & 12.291444 & 3100 & 50 & 0.003914 & 3.198785E-4 & 1.65   & 26.1    & 0.843  & Poss & Opt:M5V         \\  
   B30\_Euclid-86 & 82.84811  & 12.213705 & 3300 & 50 & 0.002927 & 8.3280065E-6& 3.575  & 24.1    & 0.825  & Poss & Opt:M4V         \\  
   B30\_Euclid-87 & 82.87079  & 12.308184 & 3000 & 50 & 0.005169 & 1.8572642E-5& 1.1    & 922.9   & 0.941  & Poss & Opt:M5V         \\  
   B30\_Euclid-89 & 82.8795   & 12.106449 & 3200 & 50 & 0.002112 & 7.819861E-6 & 5.0    & 10.3    & 0.824  & Poss & Opt:M5V         \\  
   B30\_Euclid-90 & 82.88508  & 12.070616 & 3200 & 50 & 0.003119 & 9.052931E-6 & 3.575  & 34.7    & 0.873  & Poss & Opt:M5V         \\  
   B30\_Euclid-92 & 82.93842  & 12.227889 & 3400 & 50 & 0.009472 & 5.5667224E-5& 1.65   & 332.2   & 0.873  & Poss & Opt:M4V         \\  
   B30\_Euclid-93 & 82.938484 & 12.290515 & 2900 & 50 & 0.001655 & 9.618348E-6 & 2.75   & 40.9    & 0.884  & Poss & Opt:M5V         \\  
   B30\_Euclid-94 & 82.958694 & 12.33514  & 2800 & 50 & 0.002033 & 8.383766E-6 & 2.2    & 49.5    & 0.849  & Poss & Opt:M5V         \\  
  \hline
\end{tabular}
\label{TAB_VOSA_HRD_Yes}
$\,$ \\
\end{table*}

\clearpage

%% file: 0_Tables/T_A01_B30_HB_DR3_core.izYHKsSpitzerPM_sel_HB.tex
\tiny
\begin{table*} 
\tiny
\caption{Barnard 30 possible candidate members only based on Euclid deep photometry  (sel\#1).   }
\begin{tabular}{l l l l l l l l l l l l l}
\hline
  \multicolumn{1}{l}{Name}     &
  \multicolumn{1}{c}{Object}   &
  \multicolumn{1}{c}{RA (ICRS)} &
  \multicolumn{1}{c}{DEC (ICRS)} &
  \multicolumn{1}{c}{I$_E$}    &
  \multicolumn{1}{c}{e\_I$_E$} &
  \multicolumn{1}{c}{Y$_E$}    &
  \multicolumn{1}{c}{e\_Y$_E$} &
  \multicolumn{1}{c}{J$_E$}    &
  \multicolumn{1}{c}{e\_J$_E$} &
  \multicolumn{1}{c}{H$_E$}    &
  \multicolumn{1}{c}{e\_H$_E$} &
  \multicolumn{1}{c}{Other ID$^{*}$}\\
  \multicolumn{1}{l}{} &
  \multicolumn{1}{c}{} &
  \multicolumn{1}{c}{(deg)} &
  \multicolumn{1}{c}{(deg)} &
  \multicolumn{1}{c}{(mag)} &
  \multicolumn{1}{c}{(mag)} &
  \multicolumn{1}{c}{(mag)} &
  \multicolumn{1}{c}{(mag)} &
  \multicolumn{1}{c}{(mag)} &
  \multicolumn{1}{c}{(mag)} &
  \multicolumn{1}{c}{(mag)} &
  \multicolumn{1}{c}{(mag)} &
    \multicolumn{1}{c}{}   \\
\hline
B30-Euclid-01         & J053108.18+120505.0 & 82.784104 & 12.084713 & 23.688 & 0.009 & 20.632 & 0.013 & 19.555 & 0.005 & 18.821 & 0.007  & B30-LB28a\\
B30-Euclid-02         & J053108.30+120340.5 & 82.784570 & 12.061252 & 21.031 & 0.005 & 18.821 & 0.004 & 18.297 & 0.005 & 17.889 & 0.003  &          \\
B30-Euclid-03         & J053109.50+120701.0 & 82.789584 & 12.116933 & 22.072 & 0.006 & 19.365 & 0.005 & 18.542 & 0.006 & 17.949 & 0.004  & B30-LB17b\\
B30-Euclid-04         & J053112.93+120907.0 & 82.803886 & 12.151958 & 19.944 & 0.004 & 18.131 & 0.005 & 17.673 & 0.006 & 17.404 & 0.003  &          \\
B30-Euclid-05         & J053112.97+120504.3 & 82.804030 & 12.084534 & 21.494 & 0.004 & 19.051 & 0.007 & 18.357 & 0.006 & 17.742 & 0.004  &          \\
B30-Euclid-06         & J053113.35+121025.5 & 82.805641 & 12.173745 & 22.021 & 0.005 & 19.690 & 0.009 & 19.014 & 0.004 & 18.745 & 0.009  &          \\
B30-Euclid-07         & J053113.42+121033.1 & 82.805918 & 12.175860 & 24.236 & 0.016 & 21.557 & 0.018 & 20.783 & 0.010 & 20.264 & 0.008  &          \\
B30-Euclid-08         & J053114.02+120327.8 & 82.808396 & 12.057712 & 22.392 & 0.005 & 19.494 & 0.005 & 18.525 & 0.003 & 17.908 & 0.003  & B30-LB30e\\
B30-Euclid-09         & J053114.40+120547.0 & 82.810018 & 12.096402 & 21.774 & 0.004 & 18.858 & 0.006 & 17.797 & 0.004 & 17.164 & 0.005  & B30-LB27a\\
B30-Euclid-10         & J053114.58+120814.7 & 82.810745 & 12.137424 & 21.634 & 0.005 & 18.891 & 0.005 & 18.683 & 0.009 & 18.376 & 0.005  &          \\
B30-Euclid-11         & J053118.12+120513.6 & 82.825487 & 12.087104 & 22.596 & 0.006 & 19.695 & 0.008 & 18.666 & 0.005 & 18.068 & 0.003  &          \\
B30-Euclid-12         & J053118.17+120606.3 & 82.825726 & 12.101757 & 23.398 & 0.009 & 20.497 & 0.008 & 19.827 & 0.007 & 19.090 & 0.006  &          \\
B30-Euclid-13         & J053118.39+120456.5 & 82.826643 & 12.082364 & 22.855 & 0.007 & 20.121 & 0.006 & 19.309 & 0.006 & 18.796 & 0.006  &          \\
B30-Euclid-14$^{\Pi}$ & J053119.30+120448.1 & 82.830400 & 12.080015 & 19.623 & 0.004 & 17.706 & 0.003 & 17.105 & 0.003 & 16.693 & 0.003  &          \\ 
B30-Euclid-15         & J053120.06+120536.5 & 82.833601 & 12.093461 & 22.206 & 0.007 & 19.805 & 0.010 & 19.105 & 0.006 & 18.587 & 0.005  &          \\
B30-Euclid-16         & J053121.12+120952.9 & 82.837979 & 12.164703 & 24.349 & 0.014 & 21.610 & 0.013 & 20.648 & 0.008 & 19.926 & 0.006  &          \\
B30-Euclid-17         & J053121.36+120947.8 & 82.838981 & 12.163276 & 23.451 & 0.008 & 20.507 & 0.011 & 19.262 & 0.008 & 18.497 & 0.007  &          \\
B30-Euclid-18         & J053121.46+120535.0 & 82.839426 & 12.093059 & 20.997 & 0.004 & 18.640 & 0.007 & 17.761 & 0.006 & 17.232 & 0.007  &          \\
B30-Euclid-19         & J053121.66+120915.6 & 82.840243 & 12.154336 & 23.551 & 0.010 & 20.459 & 0.012 & 19.405 & 0.009 & 18.568 & 0.007  &          \\
B30-Euclid-20         & J053122.74+120646.4 & 82.844769 & 12.112886 & 21.964 & 0.005 & 19.504 & 0.007 & 18.841 & 0.006 & 18.317 & 0.005  &          \\
B30-Euclid-21         & J053123.26+120420.7 & 82.846911 & 12.072429 & 21.170 & 0.005 & 19.154 & 0.007 & 18.376 & 0.003 & 17.999 & 0.003  &          \\
B30-Euclid-22         & J053123.48+121105.7 & 82.847813 & 12.184915 & 22.699 & 0.009 & 19.991 & 0.009 & 18.877 & 0.007 & 18.202 & 0.006  &          \\
B30-Euclid-23         & J053123.77+121142.5 & 82.849043 & 12.195134 & 22.187 & 0.006 & 19.740 & 0.010 & 18.921 & 0.007 & 18.228 & 0.006  &          \\
B30-Euclid-24$^{\Pi}$ & J053123.81+121918.2 & 82.849224 & 12.321717 & 20.413 & 0.003 & 18.603 & 0.003 & 18.020 & 0.005 & 17.731 & 0.006  &          \\ 
B30-Euclid-25         & J053124.87+120608.8 & 82.853606 & 12.102432 & 22.687 & 0.007 & 19.304 & 0.005 & 18.882 & 0.007 & 18.499 & 0.005  &          \\
B30-Euclid-26         & J053124.92+121132.9 & 82.853824 & 12.192480 & 23.922 & 0.011 & 21.280 & 0.025 & 20.174 & 0.010 & 19.549 & 0.009  &          \\
B30-Euclid-27         & J053125.10+121113.4 & 82.854597 & 12.187048 & 22.987 & 0.010 & 19.814 & 0.009 & 18.512 & 0.006 & 17.592 & 0.008  &          \\
B30-Euclid-28         & J053127.42+121803.4 & 82.864255 & 12.300935 & 23.237 & 0.010 & 20.490 & 0.008 & 19.902 & 0.005 & 19.456 & 0.009  &          \\
B30-Euclid-29         & J053127.54+120836.8 & 82.864741 & 12.143559 & 25.626 & 0.051 & 22.965 & 0.031 & 22.033 & 0.023 & 21.331 & 0.014  &          \\
B30-Euclid-30         & J053128.29+121617.7 & 82.867861 & 12.271597 & 23.590 & 0.015 & 20.873 & 0.012 & 20.055 & 0.009 & 19.458 & 0.004  &          \\
B30-Euclid-31         & J053128.40+121205.4 & 82.868328 & 12.201486 & 25.196 & 0.032 & 21.925 & 0.017 & 20.728 & 0.010 & 19.888 & 0.011  &          \\
B30-Euclid-32         & J053128.52+120740.6 & 82.868813 & 12.127931 & 24.255 & 0.015 & 21.709 & 0.018 & 20.707 & 0.014 & 20.045 & 0.012  &          \\
B30-Euclid-33         & J053128.58+120604.9 & 82.869078 & 12.101358 & 24.028 & 0.016 & 20.803 & 0.011 & 19.472 & 0.006 & 18.543 & 0.005  &          \\
B30-Euclid-34         & J053128.92+121612.9 & 82.870504 & 12.270249 & 24.582 & 0.018 & 21.779 & 0.020 & 20.903 & 0.016 & 20.358 & 0.008  &          \\
B30-Euclid-35         & J053129.10+120606.8 & 82.871259 & 12.101880 & 23.644 & 0.012 & 20.171 & 0.011 & 18.688 & 0.007 & 17.827 & 0.004  &          \\
B30-Euclid-36         & J053129.20+120815.2 & 82.871672 & 12.137554 & 21.687 & 0.004 & 19.401 & 0.010 & 18.498 & 0.005 & 17.896 & 0.004  &          \\
B30-Euclid-37         & J053129.75+120459.7 & 82.873967 & 12.083255 & 25.268 & 0.044 & 22.383 & 0.027 & 21.526 & 0.020 & 21.012 & 0.011  &          \\
B30-Euclid-38         & J053129.78+121619.8 & 82.874100 & 12.272169 & 23.149 & 0.008 & 20.543 & 0.016 & 19.429 & 0.008 & 18.846 & 0.007  &          \\
B30-Euclid-39         & J053129.95+121451.7 & 82.874801 & 12.247682 & 22.628 & 0.008 & 20.310 & 0.004 & 19.454 & 0.006 & 18.998 & 0.006  & B30-LB05c\\
B30-Euclid-40         & J053130.03+120850.1 & 82.875112 & 12.147262 & 21.901 & 0.008 & 19.675 & 0.014 & 18.996 & 0.006 & 18.559 & 0.007  &          \\
B30-Euclid-41         & J053130.14+121035.0 & 82.875568 & 12.176378 & 24.395 & 0.020 & 21.304 & 0.014 & 20.108 & 0.010 & 19.273 & 0.007  &          \\
B30-Euclid-42         & J053130.19+120718.7 & 82.875808 & 12.121860 & 25.269 & 0.038 & 22.330 & 0.038 & 21.437 & 0.015 & 20.773 & 0.015  &          \\
B30-Euclid-43         & J053130.32+121650.9 & 82.876335 & 12.280811 & 23.634 & 0.010 & 21.102 & 0.014 & 20.189 & 0.007 & 19.687 & 0.007  &          \\
B30-Euclid-44         & J053130.60+121734.6 & 82.877495 & 12.292957 & 23.183 & 0.009 & 19.926 & 0.011 & 18.678 & 0.006 & 17.593 & 0.003  & B30-LB03i\\
B30-Euclid-45         & J053130.79+121736.3 & 82.878301 & 12.293422 & 23.686 & 0.010 & 20.862 & 0.012 & 19.714 & 0.006 & 18.958 & 0.008  &          \\
B30-Euclid-46         & J053130.87+121111.6 & 82.878626 & 12.186555 & 23.862 & 0.014 & 21.043 & 0.012 & 20.221 & 0.008 & 19.620 & 0.006  &          \\
B30-Euclid-47         & J053130.96+121513.1 & 82.878984 & 12.253628 & 25.455 & 0.035 & 22.650 & 0.024 & 21.887 & 0.020 & 21.475 & 0.016  &          \\
B30-Euclid-48         & J053131.14+120645.9 & 82.879742 & 12.112760 & 22.360 & 0.007 & 19.894 & 0.008 & 18.984 & 0.006 & 18.290 & 0.006  &          \\
B30-Euclid-49         & J053131.16+120635.6 & 82.879825 & 12.109877 & 24.396 & 0.018 & 21.367 & 0.012 & 20.322 & 0.012 & 19.365 & 0.006  &          \\
B30-Euclid-50         & J053131.40+120538.5 & 82.880828 & 12.094015 & 21.620 & 0.003 & 19.481 & 0.008 & 18.760 & 0.004 & 18.263 & 0.007  &          \\
B30-Euclid-51         & J053131.44+121456.9 & 82.881017 & 12.249152 & 24.338 & 0.021 & 21.793 & 0.026 & 20.834 & 0.013 & 20.331 & 0.008  &          \\
B30-Euclid-52         & J053131.63+121450.8 & 82.881782 & 12.247449 & 22.291 & 0.008 & 19.569 & 0.007 & 18.959 & 0.007 & 18.399 & 0.007  &          \\
B30-Euclid-53         & J053131.70+121306.9 & 82.882083 & 12.218594 & 26.051 & 0.092 & 22.301 & 0.033 & 20.893 & 0.016 & 19.737 & 0.012  &          \\
B30-Euclid-54         & J053131.91+121943.5 & 82.882973 & 12.328752 & 24.712 & 0.018 & 21.716 & 0.022 & 21.022 & 0.023 & 20.304 & 0.017  &          \\
B30-Euclid-55         & J053132.01+120408.3 & 82.883356 & 12.068980 & 22.371 & 0.006 & 19.996 & 0.007 & 19.112 & 0.007 & 18.536 & 0.004  &          \\
B30-Euclid-56         & J053132.01+121308.6 & 82.883369 & 12.219047 & 25.394 & 0.050 & 22.135 & 0.026 & 20.556 & 0.017 & 19.553 & 0.008  &          \\
B30-Euclid-57         & J053132.47+120559.4 & 82.885311 & 12.099824 & 24.533 & 0.018 & 21.795 & 0.019 & 20.746 & 0.015 & 20.146 & 0.010  &          \\
B30-Euclid-58         & J053132.69+120546.7 & 82.886197 & 12.096300 & 22.394 & 0.006 & 19.625 & 0.015 & 18.649 & 0.005 & 17.958 & 0.003  &          \\
B30-Euclid-59         & J053133.56+121757.5 & 82.889815 & 12.299300 & 20.678 & 0.004 & 18.424 & 0.006 & 17.517 & 0.006 & 16.707 & 0.003  &          \\
B30-Euclid-60         & J053133.79+120625.4 & 82.890802 & 12.107047 & 24.419 & 0.019 & 21.565 & 0.022 & 20.270 & 0.007 & 19.503 & 0.007  & B30-LB18e\\
B30-Euclid-61         & J053133.96+120560.0 & 82.891498 & 12.099991 & 24.115 & 0.015 & 21.118 & 0.015 & 20.150 & 0.008 & 19.460 & 0.007  &          \\
B30-Euclid-62         & J053135.48+121812.8 & 82.897839 & 12.303553 & 25.041 & 0.027 & 22.201 & 0.025 & 21.560 & 0.024 & 20.993 & 0.014  &          \\
B30-Euclid-63         & J053135.87+120619.9 & 82.899464 & 12.105528 & 24.316 & 0.018 & 21.887 & 0.014 & 20.982 & 0.012 & 20.712 & 0.014  &          \\
B30-Euclid-64         & J053136.02+120629.7 & 82.900088 & 12.108244 & 24.926 & 0.026 & 22.108 & 0.019 & 21.094 & 0.022 & 20.416 & 0.009  &          \\
B30-Euclid-65         & J053136.48+120455.0 & 82.901999 & 12.081943 & 23.413 & 0.009 & 20.554 & 0.014 & 19.720 & 0.007 & 19.364 & 0.007  &          \\
B30-Euclid-66         & J053138.97+121008.6 & 82.912360 & 12.169068 & 24.516 & 0.018 & 21.725 & 0.018 & 21.050 & 0.016 & 20.733 & 0.009  &          \\
B30-Euclid-67         & J053140.71+121730.9 & 82.919644 & 12.291908 & 20.730 & 0.005 & 18.754 & 0.006 & 17.949 & 0.009 & 17.707 & 0.004  &          \\
B30-Euclid-68         & J053141.90+121444.2 & 82.924585 & 12.245620 & 21.028 & 0.004 & 18.928 & 0.004 & 18.422 & 0.005 & 18.242 & 0.003  &          \\
B30-Euclid-69$^{\Pi}$ & J053142.88+122006.3 & 82.928658 & 12.335093 & 19.998 & 0.004 & 18.242 & 0.005 & 17.696 & 0.006 & 17.505 & 0.0045 &          \\ 
B30-Euclid-70         & J053144.69+121953.5 & 82.936209 & 12.331533 & 24.564 & 0.016 & 21.906 & 0.015 & 20.995 & 0.013 & 20.373 & 0.009  &          \\
B30-Euclid-71$^{\Pi}$ & J053150.22+121946.1 & 82.959257 & 12.329480 & 18.991 & 0.004 & 17.224 & 0.003 & 16.975 & 0.002 & 16.697 & 0.004  &          \\ 
\hline
\end{tabular}
\label{TAB_EuclidCanSelHB}
$\,$ \\
$^{*}$ \citet{Huelamo2017-ALMA-B30,Barrado2018_B30}.\\
$^{\Pi}$ Gaia parallax with large errors.\\
\end{table*}

\clearpage

%% file: 0_Tables/T_A02_B30_HB_DR3_core.izYHKsSpitzerPM_mem_Final.tex
\tiny
\begin{table*} 
\tiny
\caption{Additional Barnard 30  possible  candidate members based on Euclid deep photometry and other ancillary data (sel\#2).   }
\begin{tabular}{l l l l l l l l l l l l}
\hline
  \multicolumn{1}{l}{Name} &
  \multicolumn{1}{c}{Object} &
  \multicolumn{1}{c}{RA (ICRS)} &
  \multicolumn{1}{c}{DEC (ICRS)} &
  \multicolumn{1}{c}{I$_E$} &
  \multicolumn{1}{c}{e\_I$_E$} &
  \multicolumn{1}{c}{Y$_E$} &
  \multicolumn{1}{c}{e\_Y$_E$} &
  \multicolumn{1}{c}{J$_E$} &
  \multicolumn{1}{c}{e\_J$_E$} &
  \multicolumn{1}{c}{H$_E$} &
  \multicolumn{1}{c}{e\_H$_E$} \\
  \multicolumn{1}{l}{} &
  \multicolumn{1}{c}{} &
  \multicolumn{1}{c}{(deg)} &
  \multicolumn{1}{c}{(deg)} &
  \multicolumn{1}{c}{(mag)} &
  \multicolumn{1}{c}{(mag)} &
  \multicolumn{1}{c}{(mag)} &
  \multicolumn{1}{c}{(mag)} &
  \multicolumn{1}{c}{(mag)} &
  \multicolumn{1}{c}{(mag)} &
  \multicolumn{1}{c}{(mag)} &
  \multicolumn{1}{c}{(mag)} \\
\hline
B30-Euclid-72$^{\Pi}$   & J053046.71+121601.8 & 82.694635 & 12.267155 & 18.392 & 0.004 & 16.834 & 0.005 & 16.490 & 0.003 & 16.270 & 0.003  \\ 
B30-Euclid-73          & J053104.73+121628.9 & 82.769717 & 12.274682 & 21.364 & 0.007 & 19.252 & 0.006 & 18.790 & 0.005 & 18.500 & 0.007  \\ 
B30-Euclid-74          & J053107.79+120618.3 & 82.782455 & 12.105072 & 22.401 & 0.009 & 19.876 & 0.011 & 19.390 & 0.007 & 18.990 & 0.007  \\ 
B30-Euclid-75$^{\Pi}$   & J053109.99+121518.0 & 82.791620 & 12.255011 & 19.652 & 0.005 & 18.206 & 0.003 & 17.890 & 0.004 & 17.670 & 0.003  \\ 
B30-Euclid-76          & J053110.34+121459.6 & 82.793095 & 12.249888 & 26.790 & 0.115 & 23.179 & 0.030 & 22.660 & 0.023 & 22.170 & 0.029  \\ 
B30-Euclid-77          & J053114.02+120325.7 & 82.808410 & 12.057149 & 24.919 & 0.023 & 21.890 & 0.018 & 20.810 & 0.012 & 20.000 & 0.009  \\ 
B30-Euclid-78          & J053114.95+121338.0 & 82.812302 & 12.227224 & 20.487 & 0.004 & 18.593 & 0.005 & 18.270 & 0.003 & 17.890 & 0.005  \\ 
B30-Euclid-79          & J053116.28+120600.8 & 82.817850 & 12.100211 & 22.635 & 0.006 & 20.489 & 0.015 & 19.600 & 0.007 & 19.100 & 0.009  \\ 
B30-Euclid-80          & J053117.07+120525.3 & 82.821130 & 12.090371 & 24.814 & 0.029 & 22.036 & 0.015 & 21.320 & 0.012 & 20.860 & 0.010  \\ 
B30-Euclid-81          & J053117.21+120533.7 & 82.821701 & 12.092693 & 24.415 & 0.015 & 21.717 & 0.019 & 20.990 & 0.012 & 20.370 & 0.010  \\ 
B30-Euclid-82          & J053119.73+121729.2 & 82.832224 & 12.291444 & 21.013 & 0.005 & 18.974 & 0.005 & 18.610 & 0.008 & 18.320 & 0.008  \\ 
B30-Euclid-83          & J053119.83+120500.1 & 82.832615 & 12.083365 & 24.770 & 0.027 & 21.942 & 0.018 & 21.060 & 0.012 & 20.370 & 0.012  \\ 
B30-Euclid-84          & J053121.16+121813.4 & 82.838166 & 12.303718 & 26.372 & 0.079 & 23.003 & 0.061 & 22.240 & 0.020 & 21.880 & 0.018  \\ 
B30-Euclid-85          & J053121.61+121111.2 & 82.840030 & 12.186455 & 26.422 & 0.100 & 23.271 & 0.038 & 22.310 & 0.024 & 21.580 & 0.018  \\ 
B30-Euclid-86          & J053123.55+121249.3 & 82.848106 & 12.213705 & 21.778 & 0.004 & 19.732 & 0.006 & 19.000 & 0.005 & 18.570 & 0.005  \\ 
B30-Euclid-87$^{\Pi}$   & J053128.99+121829.5 & 82.870792 & 12.308184 & 19.882 & 0.003 & 18.137 & 0.002 & 17.850 & 0.006 & 17.600 & 0.003  \\ 
B30-Euclid-88          & J053130.06+120911.4 & 82.875235 & 12.153165 & 24.863 & 0.025 & 21.869 & 0.017 & 21.090 & 0.012 & 20.620 & 0.014  \\ 
B30-Euclid-89          & J053131.08+120623.2 & 82.879502 & 12.106449 & 22.795 & 0.008 & 20.524 & 0.012 & 19.690 & 0.007 & 19.140 & 0.005  \\ 
B30-Euclid-90          & J053132.42+120414.2 & 82.885075 & 12.070616 & 21.707 & 0.006 & 19.593 & 0.008 & 18.930 & 0.005 & 18.510 & 0.004  \\ 
B30-Euclid-91          & J053133.85+121250.4 & 82.891029 & 12.213996 & 26.837 & 0.134 & 23.902 & 0.060 & 23.260 & 0.049 & 22.700 & 0.028  \\ 
B30-Euclid-92$^{\Pi}$   & J053145.22+121340.4 & 82.938424 & 12.227889 & 19.355 & 0.003 & 17.613 & 0.005 & 17.330 & 0.003 & 16.980 & 0.002  \\ 
B30-Euclid-93          & J053145.24+121725.9 & 82.938482 & 12.290515 & 22.162 & 0.006 & 19.889 & 0.007 & 19.260 & 0.005 & 19.090 & 0.006  \\ 
B30-Euclid-94          & J053150.09+122006.5 & 82.958695 & 12.335140 & 21.612 & 0.004 & 19.567 & 0.009 & 18.980 & 0.004 & 18.780 & 0.004  \\ 
\hline
\end{tabular}
\label{TAB_EuclidCanSel2}
$\,$ \\
$^{\Pi}$ Gaia parallax with large errors.\\
\end{table*}

\clearpage

%% file: 0_Tables/T_A03_B30_VOSA_bestfitp_mod_No.tex
\tiny
\begin{table*} 
\tiny
\caption{Rejected Barnard candidate members based on the analysis of the Spectral Distribution and the HR diagram.   }
\begin{tabular}{ l r r r r r r r r r l l } 
\hline
  \multicolumn{1}{c}{Object} &
  \multicolumn{1}{c}{RA} &
  \multicolumn{1}{c}{DEC} &
  \multicolumn{1}{c}{Teff} &
  \multicolumn{1}{c}{e\_Teff} &
  \multicolumn{1}{c}{Lbol} &
  \multicolumn{1}{c}{Lberr} &
  \multicolumn{1}{c}{Av} &
  \multicolumn{1}{c}{$\chi^2$} &
  \multicolumn{1}{c}{Fobs/Ftot} &
  \multicolumn{1}{c}{Mem?} &
  \multicolumn{1}{c}{Sp.Type}  \\ 
%
  \multicolumn{1}{c}{} &
  \multicolumn{1}{c}{(deg)} &
  \multicolumn{1}{c}{(deg)} &
  \multicolumn{1}{c}{(K)} &
  \multicolumn{1}{c}{(K)} &
  \multicolumn{1}{c}{(L$_\odot$)} &
  \multicolumn{1}{c}{(L$_\odot$)} &
  \multicolumn{1}{c}{(mag)} &
  \multicolumn{1}{c}{} &
  \multicolumn{1}{c}{} &
  \multicolumn{1}{c}{} &
  \multicolumn{1}{c}{(template)}  \\ 
\hline
   B30\_Euclid-16 & 82.83798  & 12.164703 & 3400 & 50.0 & 0.0012050397 & 7.1E-6  & 6.0   & 1.69   & 0.726 & No   & Opt:M5V         \\  
   B30\_Euclid-26 & 82.85382  & 12.19248  & 7000 & 50.0 & 0.020038817  & 6.6E-6  & 11.0  & 4.29   & 0.337 & No   & Opt:M4V         \\  
   B30\_Euclid-32 & 82.86881  & 12.127931 & 3900 & 50.0 & 0.0013184856 & 9.3E-6  & 5.5   & 20.38  & 0.682 & No   & Opt:M4V         \\  
   B30\_Euclid-34 & 82.87051  & 12.270249 & 4600 & 50.0 & 0.0013695734 & 6.8E-6  & 6.0   & 4.85   & 0.517 & No   & Opt:M4V         \\  
   B30\_Euclid-51 & 82.88102  & 12.249152 & 6500 & 50.0 & 0.0052079796 & 6.3E-6  & 9.0   & 5.15   & 0.318 & No   & Opt:M4V         \\  
   B30\_Euclid-54 & 82.88297  & 12.328752 & 3500 & 50.0 & 6.9096684E-4 & 7.1E-6  & 4.675 & 9.67   & 0.847 & No   & Opt:M8V         \\  
   B30\_Euclid-55 & 82.883354 & 12.06898  & 4700 & 50.0 & 0.008738622  & 9.1E-6  & 6.5   & 366.01 & 0.788 & No   & Opt:M4V         \\  
   B30\_Euclid-57 & 82.88531  & 12.099824 & 4600 & 50.0 & 0.0019105808 & 5.9E-6  & 7.0   & 5.3    & 0.746 & No   & Opt:M4V         \\  
   B30\_Euclid-63 & 82.89947  & 12.105528 & 3500 & 50.0 & 3.9469622E-4 & 5.0E-6  & 3.3   & 34.39  & 0.839 & No   & Opt:M6V         \\  
   B30\_Euclid-70 & 82.93621  & 12.331533 & 3300 & 50.0 & 7.4605434E-4 & 6.9E-6  & 5.5   & 2.99   & 0.749 & No   & Opt:M8V         \\  
   B30\_Euclid-72 & 82.69463  & 12.267155 & 4400 & 50.0 & 0.032047883  & 8.39E-5 & 2.475 & 383.0  & 0.924 & No   & Opt:M4V         \\  
   B30\_Euclid-78 & 82.8123   & 12.227224 & 6900 & 50.0 & 0.036090173  & 9.6E-6  & 6.0   & 482.38 & 0.513 & No   & Opt:M4V         \\  
   B30\_Euclid-80 & 82.82113  & 12.090371 & 3300 & 50.0 & 3.4439442E-4 & 3.3E-6  & 3.85  & 8.63   & 0.808 & No   & Opt:M5V         \\  
   B30\_Euclid-81 & 82.8217   & 12.092693 & 3300 & 50.0 & 7.237467E-4  & 7.5E-6  & 4.675 & 7.89   & 0.785 & No   & Opt:M7V         \\  
   B30\_Euclid-83 & 82.83262  & 12.083365 & 3700 & 50.0 & 5.9057E-4    & 6.1E-6  & 4.675 & 96.88  & 0.857 & No   & Opt:L1          \\  
   B30\_Euclid-88 & 82.87524  & 12.153165 & 3300 & 50.0 & 4.8035345E-4 & 4.7E-6  & 4.125 & 4.55   & 0.855 & No   & Opt:M4V         \\  
  \hline
\end{tabular}
\label{TAB_VOSA_HRD_No}
$\,$ \\
No SED fit for:  B30\_Euclid-18,  B30\_Euclid-29,  B30\_Euclid-37,  B30\_Euclid-42,   B30\_Euclid-47,   B30\_Euclid-62,   B30\_Euclid-76,   B30\_Euclid-77,   B30\_Euclid-84,   B30\_Euclid-85, and  B30\_Euclid-91. 
\end{table*}

\clearpage